\documentclass[9pt,aps,prd,twocolumn,superscriptaddress,showpacs]{revtex4-1}
\usepackage{amsmath,bm,amssymb,graphicx}
\usepackage{subfigure}
\usepackage[dvipdfm,colorlinks=true,citecolor=blue,pdfstartview=FitH]{hyperref}
\usepackage{longtable}
\usepackage{color,framed} 
\definecolor{shadecolor}{rgb}{1,0,0} 

\def\bea{\begin{equation}}
\def\eea{\end{equation}}
\newcommand{\rt}{Regge trajectory}
\newcommand{\rts}{Regge trajectories}
\newcommand{\sche}{Schr\"{o}dinger equation}
\newcommand{\sse}{spinless Salpeter equation}
\newcommand{\qsse}{quadratic form of the spinless Salpeter-type equation}
\newcommand{\nr}{nonrelativistic}
\newcommand{\ur}{ultrarelativistic}

\newcommand{\hspa}{\hspace{1.5mm}}
\newcommand{\bfr}{{\bf r}}
\newcommand{\bfp}{{\bf p}}
\newcommand{\bfpa}{|{\bf p}|}
\newcommand{\gev}{{\rm GeV}}

\begin{document}

\title{Structure of the meson Regge trajectories}

\author{Jiao-Kai Chen}
\email{chenjk@sxnu.edu.cn,chenjkphy@outlook.com}
\affiliation{School of Physics and Information Engineering, Shanxi Normal University, Linfen 041004, China}


\begin{abstract}
We investigate the structure of the meson Regge trajectories based on the quadratic form of the spinless Salpeter-type equation. It is found that the forms of the Regge trajectories depend on the energy region. As the employed Regge trajectory formula does not match the energy region, the fitted parameters neither have explicit physical meanings nor obey the constraints although the fitted Regge trajectory can give the satisfactory predictions if the employed formula is appropriate mathematically.
Moreover, the consistency of the Regge trajectories obtained from different approaches is discussed.
And the Regge trajectories for different mesons are presented.
Finally, we show that the masses of the constituents will come into the slope and explain why the slopes of the fitted linear Regge trajectories vary with different kinds of mesons.
\end{abstract}


\maketitle

\section{Introduction}
There are different forms of the meson {\rts} obtained from different approaches, such as the famous linear form \cite{Chew:1961ev,Chew:1962eu,Collins:1977jy,Nambu:1974zg,Polchinski:2001tt,Nielsen:2018ytt,
Olsson:1994cv,Kahana:1993yd,Lucha:1991vn,Baldicchi:1998gt,Martin:1985hw,Sonnenschein:2018fph,
Inopin:1999nf,Badalian:2019lyz,Brodsky:2016yod,Selem:2006nd,Londergan:2013dza}, the square-root form \cite{Brisudova:1999ut}, $M^2=\tau_1(J+2n_r)+\tau_2(n_r+J+1)^{-2}+\tau_0$ \cite{Sergeenko:1994ck,Sergeenko:1993sn}, $(M-m_q)^2=\pi\sigma l$ \cite{Veseli:1996gy}, $(M-2m)^2=2\pi\sigma(n+b)$ \cite{Afonin:2014nya,Afonin:2020bqc}, $M=2m+3(16\pi^2m)^{-1/3}({\sigma}l)^{2/3}$ \cite{Cotugno:2009ys,Burns:2010qq}, $M^2=a(n+b)^{\nu}$ \cite{MartinContreras:2020cyg}, $(M-m_R)^2=\beta_x(x+c_x)^{\nu}$ ($x=l,\,n_r$) \cite{chen2021} and so on. See Refs. \cite{Inopin:1999nf,Inopin:2001ub} for more discussions.
In the previous works \cite{Chen:2018hnx,Chen:2018nnr,Chen:2018bbr}, we present one new form of the meson {\rts} based on the {\qsse} (QSSE) \cite{Chen:2018hnx,Chen:2018nnr,Chen:2018bbr,Baldicchi:2007ic,Baldicchi:2007zn,Brambilla:1995bm,chenvp,chenrm}, \bea\label{nform}
M^2=\beta(c_{l}l+{\pi}n_{r}+c_0)^{2/3}+c_1,
\eea
where $l$ is the orbital angular momentum and $n_r$ is the radial quantum number. $\beta$ and $c_l$ are the universal parameters. $c_0$ and $c_1$ vary with different trajectories.
And we apply it to the heavy mesons, the heavy-light mesons and the light mesons. There are some problems remaining unclear and we attempt to resolve them in this work. For example, the fitted parameters are physically meaningful as the formula (\ref{nform}) is applied to the heavy mesons and become meaningless gradually as the quarks become lighter and lighter. Why does the slope increase as the linear formula $M^2={\beta_l}l+{\beta_{n_r}}n_r+c_0$ is employed to fit the {\rts} for the heavy-light mesons and for the heavy mesons? Is there a quantity which can distinguish the appropriate form and the inappropriate form for the given meson {\rts}?

The form of the meson {\rts} actually is complicated. For simplicity, it can be regarded as consisting of the nonlinear part corresponding to the {\nr} energy region and the linear part corresponding to the {\ur} energy region. In the intermediate region between the {\nr} region and the {\ur} region, the form of the {\rts} is not clear but expected to be nonlinear.
The structure of the meson {\rts} is discussed based on the QSSE. 

This paper is organized as follows: In Sec. \ref{sec:structure}, we present the {\rts} obtained from the QSSE in different energy regions and show the consistency of the meson {\rts} obtained from different approaches. In Sec. \ref{sec:rtms}, the {\rts} for mesons are fitted by employing the linear formula and the nonlinear formulas. In Sec. \ref{sec:beta}, we present discussions on the dependence of the slope $\beta$ on the mass of the constituents and on the string tension. The conclusions are in Sec. \ref{sec:con}.

\section{Structure of the meson {\rts}}\label{sec:structure}
In this section, we present discussions on the structure of the meson {\rts} obtained from the QSSE and on the consistency of the meson {\rts} obtained from different approaches.

\subsection{QSSE}

The {\qsse} reads \cite{Baldicchi:2007ic,Baldicchi:2007zn,Brambilla:1995bm,chenvp,chenrm}
\begin{eqnarray}\label{qsse}
M^2\Psi({\bfr})=M_0^2\Psi({\bfr})+{\mathcal U}\Psi({\bfr}),\quad M_0=\omega_1+\omega_2,
\end{eqnarray}
where $M$ is the bound state mass, $\omega_i$ is the square-root operator of the relativistic kinetic energy of constituent
\bea\label{omega}
\omega_i=\sqrt{m_i^2+{\bf p}^2}=\sqrt{m_i^2-\Delta},
\eea
\bea\label{potu}
{\mathcal U}=M_0V+VM_0+V^2.
\eea
$m_1$ and $m_2$ are the effective masses of the constituents, respectively.
For simplicity, the power-law potentials are considered,
\bea\label{potv}
V(r)={\sigma}r^{a}\,(a>0).
\eea
The confining potential is assumed to be linear with $a=1$.

\subsection{{\rts} obtained from the QSSE}

\subsubsection{{\rts}}\label{sec:subrts}
In the {\nr} limit $m_{1},m_2{\gg}{\bfpa}$, Eq. (\ref{qsse}) reduces to
\begin{eqnarray}\label{qssenr}
M^2\Psi({\bfr})&=&\left[(m_1+m_2)^2+\frac{m_1+m_2}{\mu}{\bfp}^2\right]\Psi({\bfr})\nonumber\\
&&+2(m_1+m_2)V\Psi({\bfr}),
\end{eqnarray}
where $\mu=m_1m_2/(m_1+m_2)$.
By employing the Bohr-Sommerfeld quantization approach \cite{Brau:2000st,brsom} and using Eqs. (\ref{potv}) and (\ref{qssenr}), the orbital and radial {\rts} from the QSSE have been obtained in Refs. \cite{Chen:2018nnr,Chen:2018hnx},
\begin{align}\label{rtnrs}
M^2{\sim}A_{x}x^{2a/(a+2)}\quad (x=l,\,n_r),
\end{align}
where
\begin{align}\label{qssebet}
A_l=&(2+a)(m_1+m_2)\left(\frac{\sigma^2}{(a\mu)^{a}}\right)^{1/(a+2)}\;(l{\gg}n_r),\nonumber\\
A_{n_r}=&2(m_1+m_2)\left(\frac{\sigma^2}{(2\mu)^a}\right)^{1/(a+2)} \nonumber\\
&\times\left(\frac{a\pi}{B(1/a,3/2)}\right)^{2a/(a+2)}\;(n_r{\gg}l).
\end{align}
$B(x,y)$ is the beta function \cite{betaf}. For the linear confining potential, Eq. (\ref{rtnrs}) becomes
\begin{align}\label{rtnrs1}
M^2{\sim}&3(m_1+m_2)\left(\frac{\sigma^2}{\mu}\right)^{1/3}l^{2/3},\nonumber\\
M^2{\sim}&(3\pi)^{2/3}(m_1+m_2)\left(\frac{\sigma^2}{\mu}\right)^{1/3}{n_r}^{2/3}.
\end{align}
According to Eq. (\ref{rtnrs1}), the parameterized form of the {\rts} for the {\nr} systems is suggested to be \cite{Chen:2018hnx}
\bea\label{nlrts}
M^2=\beta_x(x+c_0)^{2/3}+c_1\quad (x=l,\,n_r),
\eea
where $\beta_x$ are the universal parameters which have the theoretical values
\begin{align}\label{nrvalue}
\beta_{l}=&3(m_1+m_2)\left(\frac{\sigma^2}{\mu}\right)^{1/3},\nonumber\\
\beta_{n_r}=&(3\pi)^{2/3}(m_1+m_2)\left(\frac{\sigma^2}{\mu}\right)^{1/3},
\end{align}
respectively. $c_1$ reads
\bea\label{regc1}
c_1=(m_1+m_2)^2+{\Delta}c.
\eea
$c_0$ and ${\Delta}c$ vary with different trajectories.

In the {\ur} limit $m_1,m_2{\ll}{\bfpa}$, we obtain an auxiliary equation from Eqs. (\ref{qsse}), (\ref{omega}), (\ref{potu}) and (\ref{potv})
\bea\label{qsseu}
M^2\psi({\bf r})=4{\bf p}^2\psi({\bf r})+{\lambda}V^2\psi({\bf r})
\eea
by a very crude approximation which can also lead to the right {\rts}. In the approximation, the terms $M_0V+VM_0+V^2$ is replaced by ${\lambda}V^2$ for simplicity as $|\vec{p}|{\sim}V$. $\lambda$ is an introduced parameter. For the power-law potentials (\ref{potv}), the radial and orbital {\rts} can be obtained by employing the Bohr-Sommerfeld quantization approach,
\bea\label{urnr}
M^2{\sim}A_ll^{2a/(a+1)},\quad M^2{\sim}A_{n_r}n_r^{2a/(a+1)}
\eea
where
\begin{align}
A_l=&2^{2a/(a+1)}(\lambda\sigma^{2})^{1/(a+1)}(1+a)a^{-a/(a+1)}\;\;(l{\gg}n_r),\nonumber\\
A_{n_r}=&2^{4a/(1+a)}(\lambda\sigma^{2})^{1/(a+1)}\left[\frac{a\pi}{B(1/(2a),3/2)}\right]^{2a/(1+a)}\nonumber\\
        & (n_r{\gg}l).
\end{align}
For the linear confining potential, Eq. (\ref{urnr}) becomes
\begin{eqnarray}\label{urrt}
M^2{\sim}4{\lambda^{1/2}}{\sigma}l,{\quad} M^2{\sim}8{\lambda^{1/2}}{\sigma}n_r.
\end{eqnarray}
According to Eq. (\ref{urrt}), the parameterized {\rt} for the {\ur} systems is linear,
\bea\label{reglin}
M^2=\beta_{x}x+c_1,
\eea
where $\beta_{x}$ are universal parameters which have the theoretical values \cite{Brau:2000st}
\bea\label{urvalue}
\beta_l=8\sigma,\quad \beta_{n_r}=4\sigma\pi,
\eea
respectively. The values of ${\beta}_l$ and $\beta_{n_r}$ are different from values in Eq. (\ref{urrt}) due to the crude approximation [Eq. (\ref{qsseu})].

In the intermediate energy region where $m_1,m_2{\sim}{\bfpa}$, the square-root operator of the relativistic energy $\sqrt{m_i^2+\Delta}$ in Eq. (\ref{qsse}) cannot be expanded in the simple power series, therefore, the simple form of the {\rts} has not been obtained due to its complexity.

The ideal heavy-light systems are very special because the heavy constituent moves nonrelativistically while the light constituent moves ultrarelativistically, $m_1{\gg}{\bfpa}{\gg}m_2$. They are none of the {\nr} systems, the {\ur} systems and being in the intermediate region.
For the ideal heavy-light systems, Eq. (\ref{qsse}) reduces to
\begin{eqnarray}\label{qssehl}
M^2\Psi({\bfr})&=&\left[m_1^2+2m_1{\bfpa}+2m_1V\right]\Psi({\bfr}),
\end{eqnarray}
where the small terms have been neglected.
By employing the Bohr-Sommerfeld quantization approach \cite{Brau:2000st,brsom} and using Eqs. (\ref{potv}) and (\ref{qssehl}), the orbital and radial {\rts} from the QSSE can be obtained,
\begin{align}\label{rthlx}
M^2{\sim}A_{x}x^{a/(a+1)}\quad (x=l,\,n_r),
\end{align}
where
\begin{align}\label{rcofhl}
A_l=&2m_1(a+1){\sigma}^{1/(a+1)}\left(\frac{1}{a}\right)^{a/(a+1)}\;(l{\gg}n_r),\nonumber\\
A_{n_r}=&2m_1{\sigma}^{1/(a+1)}\left[\frac{(a+1)\pi}{a}\right]^{a/(a+1)}\;(n_r{\gg}l).
\end{align}
Fot the linear confining potential, Eq. (\ref{rthlx}) becomes
\bea\label{rcofhls}
M^2{\sim}4m_1\sqrt{\sigma}\sqrt{l},\quad M^2{\sim}2m_1\sqrt{2\pi\sigma}\sqrt{n_r}.
\eea
Some of the neglected terms in Eq. (\ref{qssehl}) can give the linear terms omitted in Eq. (\ref{rcofhls}).

Using Eq. (\ref{rcofhls}), the parameterized {\rts} for the heavy-light systems can be written as
\bea\label{reghlf}
M^2=\beta_{x}\sqrt{x+c_0}+c_1\; (x=l,n_r),
\eea
where $\beta_{x}$ are universal. They have the theoretical values \cite{Brau:2000st}
\begin{align}\label{valuehl}
\beta_{l}&=2m_1\sqrt{4\sigma},\quad \beta_{n_r}=2m_1\sqrt{2\pi\sigma},\nonumber\\
\quad c_1&=m_1^2+{\Delta}c.
\end{align}
$\Delta c$ and $c_0$ vary with different {\rts}.
Eq. (\ref{reghlf}) agrees with the form in Refs. \cite{Veseli:1996gy,Afonin:2014nya,Afonin:2020bqc,chen2021,Chen:2017fcs,Jia:2019bkr}
\bea\label{rnhlan}
(M-m_1)^2=\pi\sigma l+c_1
\eea
and with the form in Ref. \cite{Selem:2006nd}
\bea
M=m_1+\sqrt{\frac{{\sigma}l}{2}}+2^{1/4}{\kappa}{m_2}^{3/2}l^{-1/4}.
\eea

\begin{figure}[!phtb]
\centering
\includegraphics[scale=0.9]{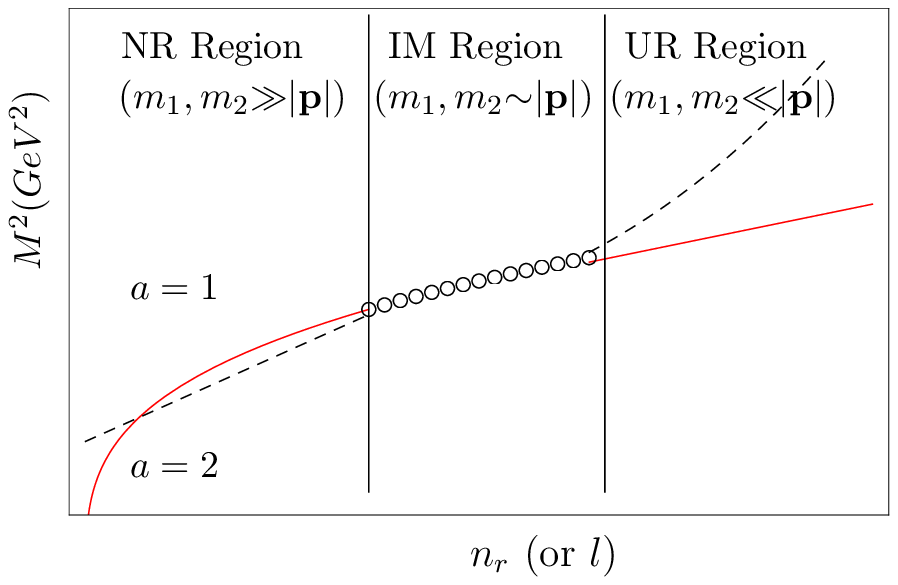}
\caption{The {\rts} in different energy regions. The NR region represents the {\nr} region, the IM region denotes the intermediate region and the UR region is the {\ur} region. The power-law potential ${\sigma}r^a$ is used for discussion. As $a=1$, the {\rt} is significantly nonlinear and concave in the NR region and is approximately linear in the UR region (the red line). As $a=2$, the {\rt} approaches linear in the  NR region and is convex in the UR region (the black dashed line). The circles represent the part of the {\rt} in the intermediate region which remains unclear.}\label{fig:struct}
\end{figure}

From the previous discussions, it is obvious that the meson {\rts} have structure and the expressions of them are complicated. If the {\rts} for the mesons in the intermediate region can be approximated by a simple power function, we have from Eqs. (\ref{rtnrs}) and (\ref{urnr})
\begin{eqnarray}\label{rtcom}
M^2{\sim}l^{\nu},\,n_r^{\nu},\quad
\left\{\begin{array}{cc}
\nu=\frac{2a}{a+2}, & \text{NR region}, \\
\frac{2a}{a+2}<\nu<\frac{2a}{a+1}, & \text{IM region},\\
\nu=\frac{2a}{a+1}, & \text{UR region}.
\end{array}\right.
\end{eqnarray}
For the linear confining potential, Eq. (\ref{rtcom}) becomes
\begin{eqnarray}\label{rtcomb}
M^2{\sim}l^{\nu},\,n_r^{\nu},\quad
\left\{\begin{array}{cc}
\nu=\frac{2}{3}, & \text{NR region}, \\
\frac{2}{3}<\nu<1, & \text{IM region},\\
\nu=1, & \text{UR region}.
\end{array}\right.
\end{eqnarray}
In Ref. \cite{MartinContreras:2020cyg}, the authors associate the index $\nu$ with the average constituent quark mass. From Eq. (\ref{rtcomb}), we can see that the meson {\rts} are concave for the linear confining potential \cite{Chen:2018bbr}.

As shown in Fig. \ref{fig:struct} and in Eq. (\ref{rtcomb}), in the {\nr} (NR) region, the {\rts} are significantly nonlinear and can be well approximated by Eq. (\ref{nlrts}), see more details in Ref. \cite{Chen:2018hnx}. In the {\ur} (UR) region, it is well-known that the {\rts} become approximately linear. In the intermediate (IM) region between the {\nr} region and the {\ur} region, the {\rts} remain unclear and are expected to be nonlinear. In one word, the approximated {\rts} range from the nonlinear form with the exponent $2/3$ to the linear form as the energy region ranges from the {\nr} region to the {\ur} region.

The proportion between the {\nr} region and the {\ur} region varies with the mesons. For the heavy mesons which are regarded as the {\nr} systems, the {\rts} take the form in Eq. (\ref{nlrts}) for small $l$ or $n_r$ and they are expected to approximate the linearity for very very large $l$ and $n_r$, see \ref{subsec:rthmeson}. For the light mesons which are taken as the relativistic systems, the {\rts} are linear approximately except for the first few points, see \ref{subsec:rtlm}. The heavy-light mesons are special and are usually regarded as the ideal heavy-light systems, see \ref{subsec:rthlm}.

\subsubsection{Discussions}\label{subsec:xidisc}
The mass of a meson can be written as
\bea\label{bmme}
M=m_1+m_2+E,
\eea
where $E$ is the interaction energy.
In the {\nr} limit, the interaction energy $E{\approx}t_{1n}n_r^{2/3},\,t_{1l}l^{2/3}$ where $t_{1n}$ and $t_{1l}$ are coefficients, see Eqs. ({\ref{nlrts}}) and (\ref{schreg32}).
Using Eq. (\ref{bmme}), we have
\bea\label{dismnr}
M^2=2t_{1x}(m_1+m_2)x^{2/3}+(m_1+m_2)^2+t_{1x}^{2}x^{4/3}\;\;(x=n_r,l).
\eea
Comparing Eqs. ({\ref{nlrts}}) and (\ref{dismnr}), we have
\begin{align}\label{nrconst}
M^2-c_1=&\beta_x(x+c_0)^{2/3}{\ll}2c_1,\nonumber\\
 c_1=&(m_1+m_2)^2+t_{1x}^{2}x^{4/3}\ge0,
\end{align}
$c_1$ in Eq. (\ref{nrconst}) is the modified form in Eq. (\ref{regc1}). The formulas in Eq. (\ref{nrconst}) are two constraints on the {\rts}.
In the {\ur} limit, $E{\approx}t_{2n}\sqrt{n_r},\,t_{2l}\sqrt{l}$, see Eq. (\ref{reglin}) and Refs. \cite{Lucha:1991vn,Brau:2000st}. Using Eq. (\ref{bmme}), we have
\bea\label{dismur}
M^2=t_{2x}^2x+2t_{2x}(m_1+m_2)\sqrt{x}+(m_1+m_2)^2\;\;(x=n_r,l).
\eea
Comparing Eqs. (\ref{reglin}) and (\ref{dismur}), we have
\begin{align}\label{urconst}
M^2-c_1=&\beta_x x{\gg}0.5c_1, \nonumber\\
c_1=&2t_{2x}(m_1+m_2)\sqrt{x}+(m_1+m_2)^2\ge0.
\end{align}
$c_1$ in Eq. (\ref{urconst}) is the modified form in Eq. (\ref{regc1}). If the constraints in (\ref{urconst}) are not obeyed, Eq. (\ref{reglin}) maybe is not appropriate again.

For the ideal heavy-light systems, $E{\sim}t_{3n}\sqrt{n_r},$ $t_{3l}\sqrt{l}$, see Eqs. (\ref{rcofhls}) and (\ref{rnhlan}).
Using Eq. (\ref{bmme}), we have
\bea\label{hlsy}
M^2=t_{3x}^2x+2t_{3x}(m_1+m_2)\sqrt{x}+(m_1+m_2)^2\;\;(x=n_r,l).
\eea
The first term on the right side of Eq. (\ref{hlsy}) is very small compared with the second term or the third term, then we have the following formulas for the ideal heavy-light systems from Eqs. (\ref{reghlf}) and (\ref{hlsy})
\begin{align}\label{hlconst}
M^2-c_1=&\beta_{x}\sqrt{x+c_0}{\ll}2c_1, \nonumber\\
c_1=&(m_1+m_2)^2+t_{3x}^2x\ge0.
\end{align}
For the common heavy-light systems, the first term in Eq. (\ref{hlsy}), $t_{3x}^2x$, is comparable with the second term and cannot be neglected, then Eq. (\ref{hlconst}) does not hold. For the common heavy-light systems, the Regge-like form (\ref{rnhlan}) \cite{Veseli:1996gy,Afonin:2014nya,Afonin:2020bqc,chen2021,Chen:2017fcs,Jia:2019bkr} will be better than the simple form (\ref{rtcomb}).

According to Eqs. (\ref{nrconst}), (\ref{urconst}) and (\ref{hlconst}), we define one quantity
\bea\label{delquant}
\xi=\frac{M^2-c_1}{c_1}.
\eea
Then we have
\begin{eqnarray}\label{quantdelt}
c_1{\ge}0,\quad \xi\left\{\begin{array}{cc}
\ll2, & \text{NR region or HLS}, \\
\gg0.5, & \text{UR region},
\end{array}\right.
\end{eqnarray}
where HLS denotes the ideal heavy-light systems. The quantity $\xi$ can be used to show the relation between the masses of the constituents and the interaction energy, similar to Eq. (\ref{fzeta}).

Physically, the nonlinear formula (\ref{nlrts}) is obtained in the {\nr} limit. Eq. (\ref{nlrts}) will be inappropriate for the {\ur} region and the parameters $\beta_x$ and $c_1$ become physically unacceptable, see Tables \ref{tab:radc}, \ref{tab:orbc}, Eqs. (\ref{nrvalue}) and (\ref{regc1}).
In practice, there are limited numbers of points on one {\rt}. Mathematically, the parameterized formula $M^2=\beta_x(x+c_0)^{2/3}+c_1$ where $x=l,\,n_r$ can fit one straight line very well like the linear formula $M^2={\beta'}x+c$ on a finite interval if $c_0$ is large. Although $c_1$ maybe is negative, the extrapolated data can be good. As $c_0$ is large,
\bea\label{nlapp}
M^2=\beta(x+c_0)^{2/3}+c_1= \frac{2}{3}\frac{\beta}{c_0^{1/3}}x + {\beta}c_0^{2/3}+c_1+\cdots.
\eea
There is a relation $\beta'{\approx}2\beta/(3c_0^{1/3})$. The nonlinear form can be used to fit not only the heavy mesons but also the light mesons, and can give the reasonable predictions, see \ref{subsec:rtlm}. The discussions on the nonlinear form in (\ref{reghlf}) and (\ref{rnhlan}) can be made similarly.

\subsection{Consistency of the meson {\rts}}\label{sub:consist}
\subsubsection{Nonrelativistic limit}\label{subsec:sch}
The {\nr} {\sche} with the power law potentials reads
\bea\label{schr}
E\psi({\bf r})=\frac{{\bf p}^2}{2\mu}\psi({\bf r})+{\sigma}r^a\psi({\bf r}),\quad (\sigma,a>0),
\eea
where $\mu=m_1m_2/(m_1+m_2)$. The {\rts} obtained from Eq. (\ref{schr}) read \cite{Brau:2000st,FabreDeLaRipelle:1988zr,Quigg:1979vr,Hall:1984wk}
\begin{align}\label{sceap}
E\sim &\left[\frac{\sigma^2}{(2\mu)^a}\right]^{1/(a+2)}\left[\frac{a\pi}{B(1/a,3/2)}\right]^{2a/(a+2)}
n_r^{2a/(a+2)},\nonumber\\
E\sim &\frac{2+a}{2}\left[\frac{\sigma^2}{(2\mu)^a}\right]^{1/(a+2)}\left(\frac{2}{a}\right)^{a/(a+2)}
l^{2a/(a+2)}.
\end{align}
Although Eq. (\ref{sceap}) is obtained in the limit $l,\,n_r\gg1$, Eq. (\ref{sceap}) is also appropriate in case of small $l$ or $n_r$ \cite{FabreDeLaRipelle:1988zr,Quigg:1979vr,Hall:1984wk,Burns:2010qq,Cotugno:2009ys}.

As $E/(m_1+m_2)\gg 1$, the following relations are obtained from Eqs. (\ref{bmme}) and (\ref{sceap}),
\begin{eqnarray}\label{rsch}
M^2{\sim}E^2{\sim}n_r^{{4a}/(a+2)},\,l^{{4a}/(a+2)}.
\end{eqnarray}
Eq. (\ref{rsch}) is the usually mentioned form as the {\rts} from the {\sche} are discussed. According to Eq. (\ref{rsch}), the {\sche} with the linear potential produces the {\rts} $M^2{\sim}$ $l^{4/3}$, $n_r^{4/3}$ which disagree with the experimental data, see Figs. \ref{fig:rthv}, \ref{fig:rthl} and \ref{fig:rtlight}.
As $E/(m_1+m_2)\ll 1$, we obtain from Eqs. (\ref{bmme}) and (\ref{sceap})
\begin{eqnarray}\label{schregconc}
M^2{\sim}E{\sim} n_r^{{2a}/(a+2)},\,l^{{2a}/(a+2)}
\end{eqnarray}
by neglecting term $E^2$. For the linear confining potential $a=1$, Eq. (\ref{schregconc}) leads to
\begin{eqnarray}\label{schreg32}
M^2\sim n_r^{{2}/{3}},\, l^{{2}/{3}},
\end{eqnarray}
Eq. (\ref{schreg32}) takes the same form of the {\rts} as Eqs. (\ref{rtnrs1}) and (\ref{nlrts}).
Mathematically, Eq. (\ref{rsch}) is applicable in case of $n_r,\,l\gg1$. Physically, the {\sche} is a {\nr} equation which is not applicable in the relativistic case, therefore, it is not Eq. (\ref{rsch}) but Eq. (\ref{schregconc}) is appropriate.
Eq. (\ref{schreg32}) is in good agreement with the experimental data, see Fig. \ref{fig:rthv} and Ref. \cite{Chen:2018hnx}. The {\sche} can produce the right {\rts} for the heavy mesons.

In the {\nr} limit, the {\rts} [Eqs. (\ref{rtnrs1}) and (\ref{nlrts})] obtained from the QSSE are consistent with the {\rts} obtained from the {\sche} and from the {\sse}. They are also in agreement with the results obtained from the Holography Inspired Stringy Hadron model \cite{Sonnenschein:2018fph}, the relativistic flux tube model or the loaded flux tube model  \cite{Burns:2010qq,Cotugno:2009ys}, the holographic AdS/QCD context \cite{MartinContreras:2020cyg} and son on.

\subsubsection{Ultrarelativistic limit}

In the {\ur} limit, the QSSE will produce the linear {\rts}, see Eqs. (\ref{urrt}) and (\ref{reglin}). It is in agreement with the {\sse} \cite{Martin:1985hw,Lucha:1991vn,Brau:2000st}, the Nambu string model \cite{Nambu:1974zg}, a first principle Salpeter equation \cite{Baldicchi:1998gt}, the relativistic Thompson equation \cite{Kahana:1993yd}, the Holography Inspired Stringy Hadron model \cite{Sonnenschein:2018fph}, the relativistic flux tube model or the loaded flux tube model \cite{Selem:2006nd}, the light-front holographic QCD \cite{Brodsky:2016yod}, the stringlike model \cite{Afonin:2014nya}, the holographic AdS/QCD context \cite{MartinContreras:2020cyg,Karch:2006pv}, the holographic model within deformed AdS$_5$ space metrics \cite{FolcoCapossoli:2019imm} and so on.

Different dynamic equations or different approximations (appropriate for different energy regions) incorporating with different kinetic terms and different potentials lead to different results. The dynamic equations with {\bfp} and $r^a$ give the $x^{a/(a+1)}$ ($x=l,\,n_r$) behavior while those with $\bfp^2$ and $r^a$ give the $x^{2a/(a+2)}$  behavior \cite{Chen:2018bbr}. Combing the obtained formulas together with masses of constituents leads to different behaviors of the {\rts} for different kinds of mesons.
We illustrate that the {\rts} obtained from the QSSE and that from other approaches are consistent with each other both in the {\nr} limit and in the {\ur} limit.

\subsection{Virial theorem}\label{sec:cons}

In the {\nr} limit, by employing the generalized virial theorem \cite{chenvp,luo:1991gvt}
\bea\label{gvtheor}
\left\langle{\bf r}\cdot\frac{\partial{F}}{\partial{\bf r}}\right\rangle=
\left\langle{\bf p}\cdot\frac{\partial{F}}{\partial{\bf p}}\right\rangle
\eea
where $F=(m_1+m_2)^2+(m_1+m_2)/{\mu}{\bfp}^2+2(m_1+m_2)V-M^2$ to the QSSE (\ref{qssenr}), we have \cite{chenvp}
\bea\label{qssevirnr}
2\left\langle\frac{m_1+m_2}{\mu} {\bf p}^2 \right\rangle= a\left\langle 2(m_1+m_2){\sigma}r^a \right\rangle.
\eea
Using Eqs. (\ref{qssenr}) and (\ref{qssevirnr}), we have
\bea\label{qssenrm}
M^2=(m_1+m_2)^2+\frac{a+2}{a}\left\langle \frac{m_1+m_2}{\mu}{\bf p}^2 \right\rangle.
\eea
Using Eqs. (\ref{qssenr}), (\ref{rtnrs}), (\ref{bmme}) and (\ref{qssenrm}), we have
\bea\label{qssegf}
E{\sim}\left\langle \frac{{\bf p}^2}{\mu} \right\rangle{\sim} \left\langle {\sigma}r^a \right\rangle
{\sim}n_r^{2a/(a+2)},\,l^{2a/(a+2)}.
\eea

In the {\ur} limit, by applying the generalized virial theorem (\ref{gvtheor}) where $F=4{\bf p}^2+\lambda{\sigma}^2r^{2a}-M^2$ to Eq. (\ref{qsseu}), we have
\bea\label{qssevirur}
\left\langle 4{\bf p}^2 \right\rangle= a\left\langle {\lambda\sigma}^2r^{2a} \right\rangle.
\eea
Using Eqs. (\ref{qsseu}), (\ref{urnr}), (\ref{bmme}) and (\ref{qssevirur}), we have
\bea\label{qsseurm}
M^2=4\left\langle {\bf p}^2 \right\rangle+\left\langle \lambda{\sigma}^2r^{2a} \right\rangle
=\frac{4a+4}{a}\left\langle {\bf p}^2 \right\rangle.
\eea
Using Eqs. (\ref{urnr}), (\ref{bmme}), (\ref{qssevirur}) and (\ref{qsseurm}), we have
\bea\label{qsseug}
E^2{\sim}\left\langle {\bf p}^2 \right\rangle{\sim}\left\langle {\lambda\sigma}^2r^{2a} \right\rangle
{\sim}n_r^{2a/(a+1)},\,l^{2a/(a+1)}.
\eea

Similarly, we have for the ideal heavy-light systems by applying Eq. (\ref{gvtheor}) to Eq. (\ref{qssehl})
\bea\label{hlvirial}
\left\langle {\bfpa} \right\rangle= a\left\langle {\lambda\sigma}r^{a} \right\rangle.
\eea
Using Eqs. (\ref{qssehl}), (\ref{rthlx}), (\ref{bmme}) and (\ref{hlvirial}), we have
\bea\label{hlms}
M^2=m_1^2+2m_1\frac{a+1}{a}\left\langle {\bfpa} \right\rangle
\eea
and
\bea\label{emhlv}
E{\sim}\left\langle {\bfpa} \right\rangle{\sim}\left\langle {\sigma}r^{a} \right\rangle
{\sim}n_r^{a/(a+1)},\,l^{a/(a+1)}.
\eea

The results obtained from the QSSE are in accordance with the results obtained from the {\sche} \cite{Quigg:1979vr} and from the {\sse} \cite{Lucha:1989jf}.
From Eqs. (\ref{qssegf}), (\ref{qsseug}) and (\ref{emhlv}), we can see that the interaction energy, the kinetic energy and the potential are in the same order. Therefore, the $\zeta$ in (\ref{fzeta}) and $\xi$ in (\ref{delquant}) are reasonable and can be used to indicate the energy regions.

\section{{\rts} for mesons}\label{sec:rtms}
In this section, the energy regions of mesons are discussed and the meson {\rts} are fitted individually. Both the nonlinear formula (\ref{nlrts}) and the linear formula (\ref{reglin}) are employed. And only those {\rts} with three points or more than three points are presented.

\begin{table}[!phtb]
\centering
\caption{The quark masses (in {\gev}) used in different models. They are used in Tables \ref{tab:orbmesons} and \ref{tab:radmesons}. The relativistic quark model (RQM) \cite{Ebert:2009ub,Ebert:2009ua,Ebert:2011jc}, the Godfrey-Isgur model (GIM) \cite{Godfrey:1985xj}, the second order Bethe-Salpeter formalism (SOBSF) \cite{Baldicchi:2003jk} and the Holography Inspired Stringy Hadron model (HISHM) \cite{Sonnenschein:2018fph} are used.}\label{tab:qmass}
\begin{tabular*}{0.46\textwidth}{@{\extracolsep{\fill}}ccccc@{}}
\hline\hline
  & RQM      & GIM    & SOBSF & HISHM \\

\hline
  $m_{u,d}$     & 0.33                                    & 0.22                       & 0.01
&0.060   \\

  $m_s$         & 0.5                                     & 0.419                       & 0.2
&0.400  \\

  $m_{c}$       & 1.55                                    & 1.628                       & 1.394
&1.490    \\

  $m_{b}$       & 4.88                                    & 4.977                        & 4.763
&4.700    \\
\hline\hline
\end{tabular*}
\end{table}

Some of the fitted resonances could be qualified as a molecular meson-meson state, like most of the axial-vector resonances \cite{Roca:2005nm}, $f_2(1270)$ \cite{Molina:2008jw}, $f'(1525)$ \cite{Geng:2008gx}, $\rho_3(1690)$, $f_4(2050)$, $\rho_5(2350)$, $f_6(2510)$, $K^*_2(1430)$, $K^*_3(1780)$, $K^*_4(2045)$, $K^\ast(52380)$ \cite{YamagataSekihara:2010qk,Roca:2010tf}. Some of them are accepted as molecules. This possible dominance of the non-$q\bar{q}$ component can change the shape of the Regge trajectory, see for instance Pelaez's work \cite{Pelaez:2017sit} and references therein. In this work, we take these resonances as the $q\bar{q}$ states like Refs. \cite{Sonnenschein:2018fph,Chen:2018nnr,Chen:2018bbr,Ebert:2009ub,Zyla:2020zbs}.

\subsection{Energy region}\label{subsec:energy}
We define a quantity $\zeta$
\bea\label{fzeta}
\zeta=\frac{E}{m_1+m_2},
\eea
where $E=M-m_1-m_2$ [Eq. (\ref{bmme})]. $\zeta$ can indicate the energy region of a state. If $\zeta\gg1$, the energy region is {\ur}. If $\zeta\ll1$, the state is in the {\nr} region or is a state of the ideal heavy-light systems. $\zeta{\sim}1$ indicates the intermediate region.

When calculating $\zeta$, we employ the relativistic quark model (RQM) \cite{Ebert:2009ub,Ebert:2009ua,Ebert:2011jc}, the Godfrey-Isgur model (GIM) \cite{Godfrey:1985xj}, the second order Bethe-Salpeter formalism (SOBSF) \cite{Baldicchi:2003jk} and the Holography Inspired Stringy Hadron model (HISHM) \cite{Sonnenschein:2018fph}. The quark masses in different models are listed in Table \ref{tab:qmass}. The calculated $\zeta$ for mesons are listed in Tables \ref{tab:orbmesons} and \ref{tab:radmesons}. For $\zeta$1(RQM), $\zeta2$(GIM) and $\zeta3$(SOBSF), the used meson masses are the theoretical masses in these models, respectively. For $\zeta4$(HISHM), the used meson masses are the experimental masses in Ref. \cite{Zyla:2020zbs}.

\begin{longtable*}{clllcrrcc}
\caption{\label{tab:orbmesons} The orbitally excited meson states used in this work. $aEb$ denotes $a\times10^{b}$. ? denotes the unwell-established states. $\dag$ denotes the data calculated by using the spin-averaged centroids. $\ast$ means that the experimental masses are used.}
\\
\hline\hline
		  Traj.   & $n^{2S+1}L_J$  & $J^{PC}$       & Meson  & Mass (MeV) \cite{Zyla:2020zbs}    &  $\zeta$1(RQM)  &  $\zeta2$(GIM)  &  $\zeta3$(SOBSF) & $\zeta4$(HISHM)$^{\ast}$ \\
\hline		
		\(\pi/b\) & $1^1S_0$       & \(0^{-+}\)     & $\pi^0$          & $134.9768\pm0.0005$
& $-$7.7E$-$1  & $-$6.6E$-$1  & 2.3E+1     & 1.2E$-$1  \\

                  & $1^1P_1$       & \(1^{+-}\)     & \(b_1(1235)\)    &1229.5${\pm}$3.2
& 9.1E$-$1   &1.8E+0      &6.6E+1      &9.2E+0 \\

		          &  $1^1D_2$      & \(2^{-+}\)     & \(\pi_2(1670)\)  & $1670.6^{+2.9}_{-1.2}$
& 1.5E+0     &2.8E+0      &8.4E+1     &1.3E+1  \\

		          &  $1^1F_3$       & \(3^{+-}\)    & \(b_3(2030)\)?   & $2032\pm12$
& 1.9E+0     &3.6E+0      &9.9E+1     &1.6E+1 \\

				  &  $1^1G_4$       &  \(4^{-+}\)   & \(\pi_4(2250)\)? &$2250\pm15$
& 2.2E+0     &4.3E+0      &1.1E+2     &1.8E+1    \\
\hline

		\(\rho/a\)&  $1^3S_1$       & \(1^{--}\)     & \(\rho(770)\)   &$775.26\pm0.25$
& 1.8E$-$1  &7.5E$-$1     &4.1E+1     &5.5E+0  \\

                  & $1^3P_2$        & \(2^{++}\)    & \(a_2(1320)\)    &$1316.9\pm0.9$
& 1.0E+0    &2.0E+0       & \hspace{1.7mm}6.6E+1$^{\dag}$    &1.0E+1 \\

	              & $1^3D_3$        & \(3^{--}\)    & \(\rho_3(1690)\) &$1688.8\pm2.1$
& 1.6E+0    &2.8E+0       & \hspace{1.7mm}8.4E+1$^{\dag}$    &1.3E+1 \\

                  & $1^3F_4$        &  \(4^{++}\)   & \(a_4(1970)\)    &$1967\pm16$
& 2.1E+0     &3.6E+0     & \hspace{1.7mm}9.9E+1$^{\dag}$     &1.5E+1 \\

                  & $1^3G_5$        &  \(5^{--}\)   & \(\rho_5(2350)\)? &$2330\pm35$
& 2.4E+0     &4.2E+0     & \hspace{1.7mm}1.1E+2$^{\dag}$      &1.8E+1  \\

                  & $1^3H_6$        & \(6^{++}\)    & \(a_6(2450)\)?   &$2450\pm130$
& 2.8E+0     &           & \hspace{1.7mm}1.2E+2$^{\dag}$      &1.9E+1  \\
\hline


\(\eta^\prime/h\)      & $1^1P_1$        &  \(1^{+-}\)    & \(h_1(1170)\)   &$1166\pm6$
& 9.1E$-$1   &1.8E+0     &        & 8.7E+0 \\

                   & $1^1D_2$       & \(2^{-+}\)     & \(\eta_2(1645)\) &$1617\pm5$
& 1.5E+0     &2.8E+0    &         &1.2E+1  \\

                  &  $1^1F_3$       & \(3^{+-}\)     & \(h_3(2025)\)?  &$2025\pm20$
& 1.9E+0     &3.6E+0    &         &1.6E+1  \\

                   &  $1^1G_4$      & \(4^{-+}\)     & \(\eta_4(2330)\)? &$2328\pm38$
& 2.2E+0     &4.3E+0    &         &1.8E+1        \\
\hline

		\(\omega/f\)& $1^3S_1$     & \(1^{--}\)      & \(\omega(782)\)  &$782.65\pm0.12$
&1.8E$-$1    &7.7E$-$1   &           & 5.5E+0 \\

                   &  $1^3P_2$     & \(2^{++}\)       & \(f_2(1270)\)   &$1275.5\pm0.8$
&1.0E+0      &1.9E+0    &            &9.6E+0  \\

                    & $1^3D_3$     & \(3^{--}\)      & \(\omega_3(1670)\) &$1667\pm4$
& 1.6E+0     &2.8E+0   &             &1.3E+1 \\
							
                    & $1^3F_4$      & \(4^{++}\)     & \(f_4(2050)\)      &$2018\pm11$
& 2.1E+0     &3.6E+0   &              &1.6E+1  \\

                    & $1^3G_5$      & \(5^{--}\)     & \(\omega_5(2250)\)? &$2250\pm70$
&2.4E+0      &4.2E+0   &             &1.8E+1  \\

                    & $1^3H_6$      & \(6^{++}\)     & \(f_6(2510)\)?      &$2465\pm50$
& 2.8E+0     &         &            &  2.0E+1 \\
\hline
							
		\(K\)       & $1^1S_0$       &  \(0^-\)       & $K^0$          &$497.611\pm0.013$
&$-$4.2E$-$1     &$-$2.6E$-$1      &2.1E+0      &8.2E$-$2 \\

					& $1^1P_1$       &\(1^+\)         & \(K_1(1270)\)  &$1253\pm7$
&5.6E$-$1        &1.1E+0           &5.7E+0      &1.7E+0   \\

                    & $1^1D_2$       & \(2^-\)        & \(K_2(1770)\)  &$1773\pm8$
&1.1E+0         &1.8E+0            &7.4E+0       &2.9E+0 \\
\hline

		\(K^*\)     & $1^3S_1$        & \(1^-\)        & $K^*(892)$   &$891.66\pm0.26$
&8.1E$-$2      &4.1E$-$1             &3.5E+0      &9.4E$-$1 \\

					& $1^3P_2$        & \(2^+\)        & \(K^*_2(1430)\) &$1427.3\pm1.5$
&7.2E$-$1       &1.2E+0      & \hspace{1.7mm}5.7E+0$^{\dag}$  &2.1E+0 \\

                    & $1^3D_3$        & \(3^-\)        & \(K^*_3(1780)\) &$1776\pm7$
&1.2E+0         &1.8E+0      &\hspace{1.7mm}7.4E+0$^{\dag}$   &2.9E+0 \\

                    & $1^3F_4$        & \(4^+\)        & \(K^*_4(2045)\) &$2048^{+8}_{-9}$
&1.5E+0         &2.3E+0      &\hspace{1.7mm}8.7E+0$^{\dag}$   &3.5E+0  \\

                    &$1^3G_5$         &  \(5^-\)       & \(K^*_5(2380)\)? &$2382\pm24$
& 1.8E+0        &2.7E+0      &\hspace{1.7mm}9.9E+0$^{\dag}$    &4.2E+0 \\
\hline

	   \(\phi/f'\)  &$1^3S_1$         & \(1^{--}\)     & \(\phi(1020)\)  &$1019.461\pm0.016$
&3.8E$-$2     &2.2E$-$1    &1.6E+0   &2.7E$-$1  \\

	            	&$1^3P_2$         & \(2^{++}\)      & \(f_2^\prime(1525)\) &$1517.4\pm2.5$
&5.3E$-$1     &8.3E$-$1    &\hspace{1.7mm}2.7E+0$^{\dag}$  &9.0E$-$1  \\

                   	&$1^3D_3$          &  \(3^{--}\)    & \(\phi_3(1850)\)   &$1854\pm7$
&9.5E$-$1     &1.3E+0      &\hspace{1.7mm}3.6E+0$^{\dag}$  &1.3E+0 \\
                 	&$1^3F_4$          &  \(4^{++}\)    & \(f_4(2300)\)?   &$2320\pm60$
&1.3E+0       &1.6E+0      &\hspace{1.7mm}4.3E+0$^{\dag}$  &1.9E+0 \\
\hline
		\(D\)   &  $1^1S_0$   & \(0^-\)            & \(D^0\)            &$1864.83\pm0.05$
&$-$4.8E$-$3    & 1.7E$-$2    & 3.4E$-$1   &2.0E$-$1   \\

	         	&  $1^1P_1$     &  \(1^+\)           & \(D_1(2420)^0\)  &$2420.8\pm0.5$
&2.9E$-$1     &3.2E$-$1    & 7.6E$-$1       &5.6E$-$1  \\

             	&  $1^1D_2$    &\(2^-\)             & \(D(2740)^0\)?    &$2737\pm12$
&4.9E$-$1     &            &               &7.7E$-$1  \\
\hline						
		
		\(D^*\) & $1^3S_1$    & \(1^-\)            & \(D^*(2007)^0\)   &$2006.85\pm0.05$
&6.9E$-$2   &1.0E$-$1      & 4.4E$-$1     &2.9E$-$1   \\

	         	& $1^3P_2$   & \(2^+\)            & \(D^*_2(2460)^0\)  &$2460.7\pm0.4$
&3.1E$-$1    &3.5E$-$1     &\hspace{1.7mm}7.6E$-$1$^{\dag}$  &5.9E$-$1 \\
						
	            &  $1^3D_3$    & \(3^-\)           & \(D^*_3(2750)\)?  &$2763.5\pm3.4$
&5.2E$-$1   &5.3E$-$1      &     &7.8E$-$1  \\
\hline								
		
%

		\(D^*_s\) &  $1^3S_1$     &  \(1^-\)          & \(D^{*\pm}_s\) &$2112.2\pm0.4$
&3.0E$-$2  &4.1E$-$2      &3.3E$-$1      &1.2E$-$1  \\

	            &   $1^3P_2$    &  \(2^+\)            & \(D^*_{s2}(2573)\)  &$2569.1\pm0.8$
&2.5E$-$1  &2.7E$-$1     &\hspace{1.7mm}6.0E$-$1$^{\dag}$ &3.6E$-$1 \\

	          	&   $1^3D_3$   & \(3^-\)            & \(D^{*}_{s3}(2860)^{\pm}\)? &$2860\pm7$
&4.5E$-$1  &4.3E$-$1     &         &5.1E$-$1       \\
\hline
								
		\(\Psi\) & $1^3S_1$     & \(1^{--}\)         & \(J/\Psi(1S)\)    &$3096.900\pm0.006$
&$-$1.3E$-$3  &$-$4.8E$-$2      & 1.1E$-$1   &3.9E$-$2 \\
		
            	& $1^3P_2$       & \(2^{++}\)         & \(\chi_{c2}(1P)\) &$3556.17\pm0.07$
&1.5E$-$1   &9.0E$-$2       &\hspace{1.7mm}2.7E$-$1$^{\dag}$  &1.9E$-$1  \\

\hline							
		\(\eta_c\) & $1^1S_0$    & \(0^{-+}\)         & \(\eta_c(1S)\)  &$2983.9\pm0.5$
&$-$3.8E$-$2  &$-$8.8E$-$2      & 7.0E$-$2   &1.3E$-$3  \\
		
               	&   $1^1P_1$    & \(1^{+-}\)          &  \(h_c(1P)\)    &$3525.38\pm0.11$
&1.4E$-$1    &8.1E$-$2      &2.7E$-$1      &1.8E$-$1 \\

\hline		
		\(B\)   &   $1^1S_0$       &  \(0^-\)           & \(B^{\pm}\)   &$5279.34\pm0.12$
&1.3E$-$2    &2.2E$-$2      &1.0E$-$1    &1.1E$-$1  \\

	         	&   $1^1P_1$     & \(1^+\)             & \(B_1(5721)^+\)  &$5725.9^{+2.5}_{-2.7}$
&9.8E$-$2    &             & 2.1E$-$1    &2.0E$-$1 \\

\hline		
		\(B^*\) &   $1^3S_1$       & \(1^-\)           & \(B^*\)          &$5324.70\pm0.21$
&2.2E$-$2   &3.3E$-$2    & 1.2E$-$1      &1.2E$-$1 \\

	         	&   $1^3P_2$     & \(2^+\)             & $B^*_2(5747)^0$  & $5737.2\pm0.7$
&1.0E$-$1   &1.2E$-$1    &\hspace{1.7mm}2.1E$-$1$^{\dag}$   &2.1E$-$1  \\
\hline			
		
		\(B_s\) &   $1^1S_0$     & \(0^-\)            & \(B_s^0\)          &$5366.88\pm0.14$
&$-$1.5E$-$3  &$-$1.1E$-$3   & 8.1E$-$2   &5.2E$-$2  \\

	         	&   $1^1P_1$         & \(1^+\)        & \(B_{s1}(5830)^0\)  &$5828.70\pm0.20$
&8.4E$-$2   &            & 1.8E$-$1       &1.4E$-$1  \\

\hline
		\(B^*_s\) &   $1^3S_1$        & \(1^-\)       & \(B^*_s\)       &$5415.4^{+1.8}_{-1.5}$
&6.3E$-$3   &1.0E$-$2    & 9.4E$-$2        &6.2E$-$2  \\

                 &     $1^3P_2$     & \(2^+\)         & \(B^*_{s2}(5840)^0\) &$5839.86\pm0.12$
&8.6E$-$2   &9.0E$-$2    & \hspace{1.7mm}1.8E$-$1$^{\dag}$  &1.5E$-$1  \\
\hline
								
		\(\Upsilon\) &   $1^3S_1$    & \(1^{--}\)     & \(\Upsilon(1S)\)  &$9460.30\pm0.26$
& $-$3.1E$-$2  &$-$5.0E$-$2   &\hspace{-2.2mm}$-$6.9E$-$3  &6.4E$-$3 \\
		
            	&  $1^3P_2$       & \(2^{++}\)    & \(\chi_{b2}(1P)\) &$9912.21\pm0.26\pm0.31$
&1.6E$-$2    &$-$5.4E$-$3   & {\hspa}4.0E$-$2$^{\dag}$    &5.4E$-$2 \\

\hline
								
		\(\eta_b\) &    $1^1S_0$      & \(0^{-+}\)   & \(\eta_b(1S)\)  &$9398.7\pm2.0$
&$-$3.7E$-$2   &$-$5.6E$-$2   & \hspace{-2.2mm}$-$1.6E$-$2 & \hspace{-2.2mm}$-$1.4E$-$4 \\
		
                  &    $1^1P_1$        & \(1^{+-}\)  &  \(h_b(1P)\)   &$9899.3\pm0.8$
& 1.4E$-$2   &$-$7.4E$-$3   & 4.0E$-$2   &5.3E$-$2 \\
\hline\hline
\end{longtable*}

As shown in Tables \ref{tab:orbmesons} and \ref{tab:radmesons}, $\zeta>1$ for most of the light mesons. However, $\zeta$ is sometimes small for the first few states on a {\rt} for the light mesons. It is suggested that the first few states are neglected as fitting the {\rt} to avoid the {\nr} effect. For the light mesons, the linear form of a {\rt} is a good approximation, see Fig. \ref{fig:rtlight}.
For the heavy-light mesons, $\zeta{\sim}10^{-1}$ which implies crudely the intermediate region. The fits of the nonlinear {\rts} (\ref{reghlf}) give small or negative $c_1$ which are physically meaningless, see \ref{subsec:rthlm} and Fig. \ref{fig:rthl}.
For most of the heavy mesons, $\zeta{\sim}10^{-2}$ which indicates the {\nr} region. And the nonlinear form of a {\rt} (\ref{nlrts}) will be a good approximation, see Fig. \ref{fig:rthv}.

\begin{longtable*}{cclllrrcc}
\caption{\label{tab:radmesons} Same as Table \ref{tab:orbmesons} except for the radially excited states.}
\\
\hline\hline
		Traj.   & $n^{2S+1}L_J$     & \(I(J^{PC})\)   & Meson  & $\;\;\;\;$ Mass (MeV) \cite{Zyla:2020zbs} &  $\zeta$1(RQM)   & $\zeta$2(GIM)  & $\zeta$3(SOBSF)  & $\zeta$4(HISHM)$^\ast$ \\
\hline

		\(\pi\)   & $1^1S_0$        & \(1(0^{-+})\) & \(\pi^0\) & $134.9768\pm0.0005$
& $-$7.7E$-$1    &$-$6.6E$-$1  & 2.3E+1  &1.2E$-$1 \\

                  &  $2^1S_0$     &              & \(\pi(1300)\)& $1300\pm100$
& 9.6E$-$1   &2.0E+0           &6.5E+1    &9.8E+0 \\

		          &  $3^1S_0$       &              & \(\pi(1800)\) & $1810^{+9}_{-11}$
&1.7E+0      &3.3E+0           &9.0E+1  &1.4E+1 \\

		          &  $4^1S_0$    &               & \(\pi(2070)\)? & $2070\pm35$
&2.1E+0       &                &        & 1.6E+1  \\

			    	& $5^1S_0$     &              & \(\pi(2360)\)?	& $2360\pm25$
&2.6E+0       &                &        &1.9E+1 \\
\hline

		\(a_1\)     & $1^3P_1$   & \(1(1^{++})\) & \(a_1(1260)\) & $1230\pm40$
&9.0E$-$1      & 1.8E+0        &\hspa6.6E+1$^\dag$  &9.3E+0 \\

					& $2^3P_1$	&                 & \(a_1(1640)\) & $1655\pm16$
&1.6E+0        &3.1E+0         &          &1.3E+1 \\

					& $3^3P_1$  &                 & \(a_1(2095)\)? & $2096\pm17\pm121$
&2.1E+0        &              &           &1.6E+1 \\

					& $4^3P_1$	&                 & \(a_1(2270)\)? & $2270^{+55}_{-40}$
&2.5E+0        &              &             &1.8E+1  \\
\hline

		\(\pi_2\)  & $1^1D_2$     & \(1(2^{-+})\) & \(\pi_2(1670)\) & $1670.6^{+2.9}_{-1.2}$
&1.5E+0        &2.8E+0        &8.4E+1      &1.3E+1 \\

		     		& $2^1D_2$     &               & \(\pi_2(2005)\)? & $1963^{+17}_{-27}$
&2.0E+0         &            &             &1.5E+1  \\

					&$3^1D_2$&                     &\(\pi_2(2285)\)? & $2285\pm20\pm25$
&2.4E+0         &            &            &1.8E+1   \\
\hline

		\(h_1\)		& $1^1P_1$ &  \(0(1^{+-})\) & \(h_1(1170)\) &	$1166\pm6$
&9.1E$-$1      &1.8E+0       &             &8.7E+0  \\

					&$2^1P_1$&                   & \(h_1(1595)\)?  & $1594^{+18}_{-60}$	
&1.6E+0        &3.0E+0       &             &1.2E+1  \\
							
					&$3^1P_1$&                   & \(h_1(1965)\)?  & $1965\pm45$
&2.0E+0        &             &              &1.5E+1 \\
							
					&$4^1P_1$&                    & \(h_1(2215)\)? & $2215\pm40$
&2.4E+0        &            &               &1.7E+1 \\
\hline
							
		\(\omega\) & $1^3S_1$ & \(0(1^{--})\) & \(\omega(782)\) & $782.65\pm0.12$
&1.8E$-$1     &7.7E$-$1     &                &5.5E+0  \\

					&$2^3S_1$&                  & \(\omega(1420)\) & $1410\pm60$
&1.3E+0       &2.3E+0       &                &1.1E+1  \\

					&$3^3S_1$&                  & \(\omega(1650)\) & $1670\pm30$
&1.9E+0      &              &                &1.3E+1  \\

					&$4^3S_1$&                  & \(\omega(1960)\)? & $1960\pm25$
&2.3E+0      &               &               &1.5E+1   \\	
						
					&$5^3S_1$&                  & \(\omega(2290)\)? & $2290\pm20$
&2.8E+0      &               &               &1.8E+1   \\
\hline

		\(K\)   &  $1^1S_0$  &	\(0(0^{-})\)	& $K^0$ & $497.611\pm0.013$
&$-$4.2E$-$1   &$-$2.6E$-$1     &2.1E+0       &8.2E$-$2  \\

				   &$2^1S_0$	&	            &$K(1460)$	 & $1482.40\pm3.58\pm15.22$
&8.5E$-$1    &1.3E+0            &5.7E+0        &2.2E+0  \\

				   &$3^1S_0$	&	            &$K(1830)$?	 & $1874^{+70}_{-120}$
&1.5E+0      &2.2E+0            &8.0E+0        &3.1E+0  \\
\hline

		\(\phi\)   & $1^3S_1$  &	\(0(1^{--})\)	& \(\phi(1020)\) & $1019.461\pm0.016$
&3.8E$-$2    &2.2E$-$1         & 1.6E+0      &2.7E$-$1 \\

				   &$2^3S_1$	&	           &	\(\phi(1680)\) & $1680\pm20$
&7.0E$-$1    &1.0E+0            &3.1E+0      &1.1E+0  \\

				   &$3^3S_1$	&              & \(\phi(2170)\) & $2160\pm80$
&1.1E+0      &                  &4.2E+0      &1.7E+0  \\
\hline

		\(\eta_{c}\) & $1^1S_0$ & \(0(0^{-+})\) & \(\eta_{c}(1S)\) & $2983.9\pm0.5$
&$-$3.8E$-$2  &$-$8.8E$-$2      & 7.0E$-$2   &1.3E$-$3   \\
		
				     &	$2^1S_0$&               & \(\eta_{c}(2S)\) & $3637.5\pm1.1$
&1.7E$-$1   &1.1E$-$1           & 2.8E$-$1   &2.2E$-$1 \\
\hline
		
		\(\Psi\)     &$1^3S_1$  & \(0(1^{--})\) & \(J/\Psi(1S)\) & $3096.900\pm0.006$
&$-$1.3E$-$3  &$-$4.8E$-$2      &1.1E$-$1    &3.9E$-$2  \\

					&$2^3S_1$&               	& \(\Psi(2S)\) & $3686.10\pm0.06$
&1.9E$-$1   &1.3E$-$1           &3.1E$-$1    &2.4E$-$1   \\

					&$3^3S_1$&                  & \(\Psi(4040)\) & $4039\pm1$
&3.0E$-$1   &2.6E$-$1           &4.4E$-$1    &3.6E$-$1   \\

					&$4^3S_1$&	                & \(\Psi(4415)\) & $4421\pm4$
&4.3E$-$1   &                   &5.6E$-$1     &4.8E$-$1    \\
\hline
		
		\(\chi_{c2}\)& $1^3P_2$ & \(0(2^{++})\) & \(\chi_{c2}(1P)\) & $3556.17\pm0.07$
&1.5E$-$1   &9.0E$-$2           &{\hspa}2.7E$-$1$^\dag$   &1.9E$-$1  \\
		
					 &	$2^3P_2$&              & \(\chi_{c2}(3930)\) & $3922.2\pm1.0$
&2.7E$-$1   &2.2E$-$1           &{\hspa}4.1E$-$1$^\dag$   &3.2E$-$1  \\
\hline
		
		\(\Upsilon\)  &$1^3S_1$& \(0(1^{--})\) & \(\Upsilon(1S)\) & $9460.30\pm0.26$
&$-$3.1E$-$2  &$-$5.0E$-$2      &\hspace{-2.6mm}$-$6.9E$-$3  &6.4E$-$3 \\

					&$2^3S_1$ &                & \(\Upsilon(2S)\) & $10023.26\pm0.31$
&2.7E$-$2  &4.6E$-$3             & 5.1E$-$2   &6.6E$-$2  \\

					&$3^3S_1$ &                & \(\Upsilon(3S)\) & $10355.2\pm0.5$
&6.1E$-$2  &4.0E$-$2             &8.6E$-$2    &1.0E$-$1  \\

					&$4^3S_1$ &                & \(\Upsilon(4S)\) & $10579.4\pm1.2$
&8.5E$-$2  &6.8E$-$2             &1.1E$-$1    &1.3E$-$1  \\

					&$5^3S_1$ &         & \(\Upsilon(10860)\) & $10885.2^{+2.6}_{-1.6}$
&1.1E$-$1  &9.3E$-$2             &1.4E$-$1    &1.6E$-$1  \\

					&$6^3S_1$ &                & \(\Upsilon(11020)\)& $11000\pm4$
&1.4E$-$1  &1.2E$-$1             &1.6E$-$1    &1.7E$-$1  \\
\hline

		\(\chi_{b1}\) &$1^3P_1$ &	\(0(1^{++})\) & \(\chi_{b1}(1P)\) & $9892.78\pm0.26\pm0.31$
&1.4E$-$2  &$-$7.4E$-$3         &{\hspa}4.0E$-$2$^\dag$  &5.2E$-$2   \\

				    	&$2^3P_1$	&             & \(\chi_{b1}(2P)\) & $10255.46\pm0.22\pm0.50$
&5.1E$-$2  &3.0E$-$2            &{\hspa}7.7E$-$2$^\dag$   &9.1E$-$2  \\

						&$3^3P_1$&                & \(\chi_{b1}(3P)\) & $10513.4\pm0.7$
&8.0E$-$2  &                    &               &1.2E$-$1  \\
\hline

		\(\chi_{b2}\)  &$1^3P_2$ &	\(0(2^{++})\) & \(\chi_{b2}(1P)\) & $9912.21\pm0.26\pm0.31$
&1.6E$-$2  &$-$5.4E$-$3       &{\hspa}4.0E$-$2$^\dag$  &5.4E$-$2   \\

				    	&$2^3P_2$	&              & \(\chi_{b2}(2P)\) & $10268.65\pm0.22\pm0.50$
&5.2E$-$2  &3.1E$-$2           &{\hspa}7.7E$-$2$^\dag$   &9.2E$-$2  \\

						&$3^3P_2$&                 & \(\chi_{b2}(3P)\) &$10524.0\pm0.8$
&8.1E$-$2  &                   &                 &1.2E$-$1  \\
\hline\hline
\end{longtable*}

\subsection{{\rts} for the heavy mesons}\label{subsec:rthmeson}
For the nonlinear fit [Eq. (\ref{nlrts})], all states on the {\rts} are used. For the linear fit [Eq. (\ref{reglin})], the last four states are used if the data points on the {\rt} are equal to or greater than four.

\begin{figure*}[!phtb]
\centering
\subfigure[]{\label{fig:subfigure:bs}\includegraphics[scale=0.9]{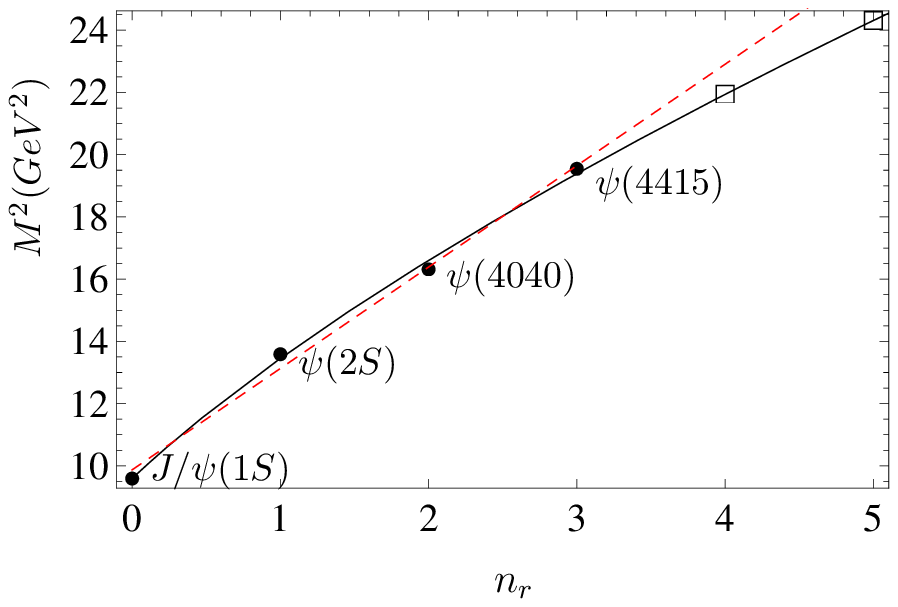}}
\subfigure[]{\label{fig:subfigure:bp}\includegraphics[scale=0.9]{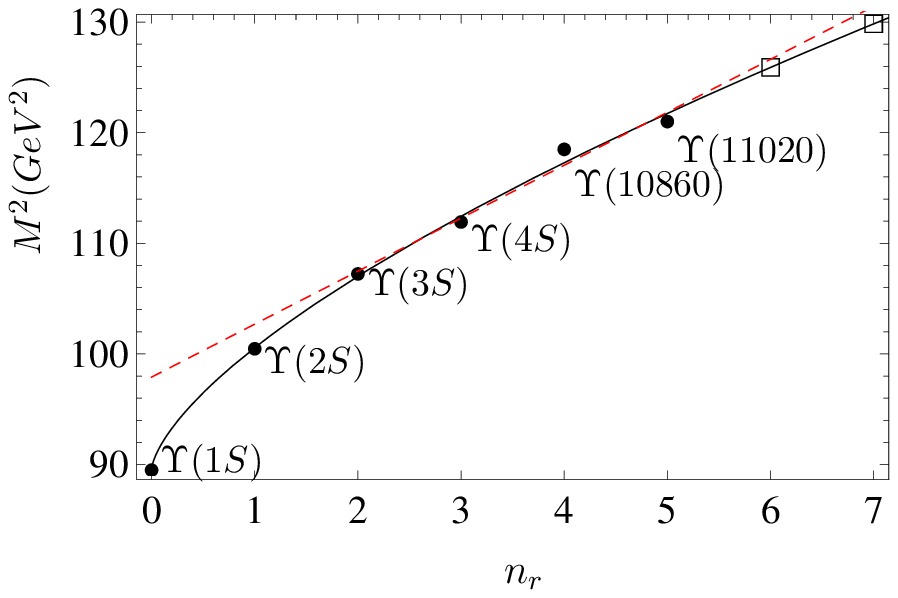}}
\subfigure[]{\label{fig:subfigure:bp}\includegraphics[scale=0.9]{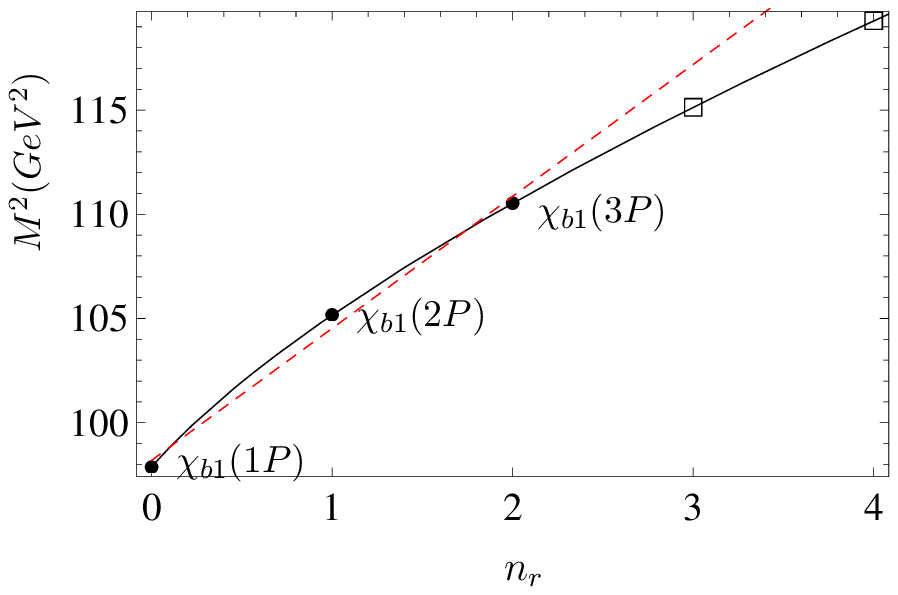}}
\subfigure[]{\label{fig:subfigure:bp}\includegraphics[scale=0.9]{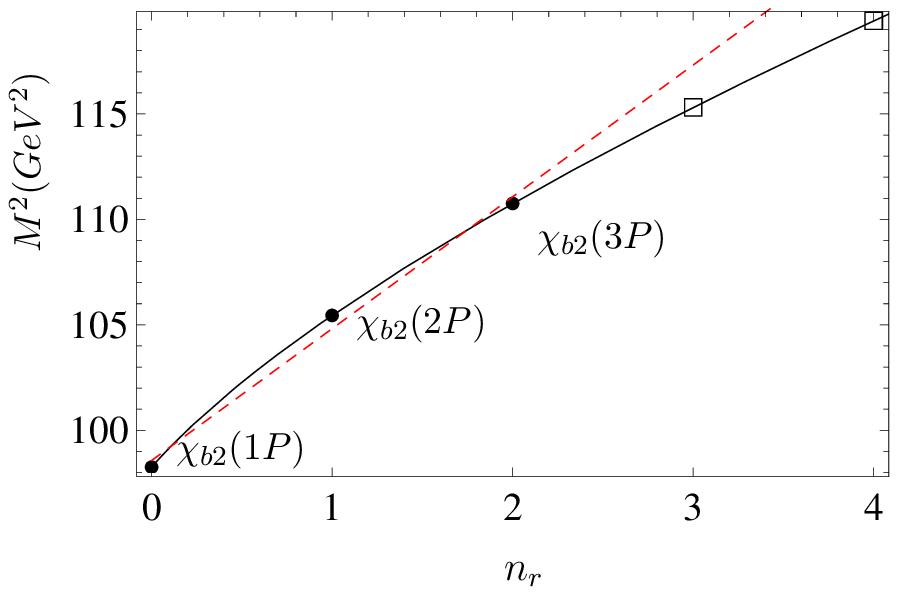}}
\caption{The {\rts} for the heavy mesons fitted by using the linear formula (\ref{reglin}) (the red dashed line) and by using the nonlinear formula (\ref{nlrts}) (the black line). The fitted {\rts} are listed in Tables \ref{tab:radc} and \ref{tab:orbc}. The well-established states are given by solid dots and the unwell-established stated are given by circles. Open squares are the predicted masses by the nonlinear formula.}\label{fig:rthv}
\end{figure*}

\begin{table*}[!phtb]
\centering
\caption{$\Upsilon(nS)$ states. The experimental masses (in {\gev}) and the predicted masses fitted by the nonlinear formula $M^2=89.55+11.01n_r^{2/3}$(Fit1) and by the linear formula $M^2=97.90+4.79n_r$(Fit2) are listed. When calculating the $\zeta1$ and $\zeta2$ [Eq. (\ref{fzeta})], $m_b=4.88$ \cite{Ebert:2009ub,Ebert:2009ua,Ebert:2011jc}, $M$ are obtained by Fit1 and Fit2, respectively.}\label{tab:botenergy}
\begin{tabular*}{\textwidth}{@{\extracolsep{\fill}}rccccccc@{}}
\hline\hline
         & Mass \cite{Zyla:2020zbs} & Fit1     &  Fit2     & $\zeta$1   & $\zeta$2         & $\xi$1   & $\xi$2 \\

\hline
  $1^3S_1$     & 9.46030   & 9.4631  &           & \hspace{-2.5mm}$-$3.0E$-$2  &    & 0.0   & 0.0 \\

  $2^3S_1$     & 10.02326  & 10.028  &           &  2.7E$-$2   &                   & 0.12   & 0.05 \\

 $3^3S_1$      & 10.3552   & 10.345  & 10.367  &  6.0E$-$2   & 6.2E$-$2          & 0.20   & 0.10 \\

 $4^3S_1$      & 10.5794   & 10.604  & 10.596   &  8.7E$-$2  & 8.6E$-$2            & 0.26   & 0.15  \\

 $5^3S_1$      & 10.8852   & 10.830   & 10.819  &  1.1E$-$1  & 1.1E$-$1            & 0.30   & 0.20 \\

 $6^3S_1$      & 11.000     & 11.034   & 11.039  &  1.3E$-$1  & 1.3E$-$1           & 0.36   & 0.24\\

 $7^3S_1$      &           & 11.221   & 11.253  &  1.5E$-$1  &  1.5E$-$1           & 0.41   & 0.29 \\

 $8^3S_1$      &           & 11.395   & 11.464  &  1.7E$-$1  &  1.7E$-$1           & 0.45   & 0.34 \\

 $9^3S_1$      &           & 11.558   & 11.671  &  1.8E$-$1  & 2.0E$-$1            & 0.49   & 0.39 \\

 $10^3S_1$      &          & 11.713   & 11.875  &  2.0E$-$1  & 2.2E$-$1            & 0.53   & 0.44 \\

 $11^3S_1$      &          & 11.860  &  12.075  &  2.2E$-$1  & 2.4E$-$1            & 0.57   & 0.49 \\

 $12^3S_1$      &          &  12.000  & 12.272  &  2.3E$-$1  & 2.6E$-$1            & 0.61   & 0.54 \\

 $13^3S_1$      &          & 12.135  &  12.465  &  2.4E$-$1  &  2.8E$-$1           & 0.64   & 0.59  \\

 $14^3S_1$      &          & 12.265  &  12.656  &  2.6E$-$1  & 3.0E$-$1            & 0.68   & 0.64   \\

 $15^3S_1$      &          & 12.390   & 12.844  &  2.7E$-$1  & 3.2E$-$1            & 0.71   & 0.68 \\
\hline\hline
\end{tabular*}
\end{table*}

As shown in Fig. \ref{fig:rthv} and Tables \ref{tab:orbmesons} and \ref{tab:radmesons}, $\zeta$ [Eq. (\ref{fzeta})] are small for the heavy mesons especially for the bottomonia which indicates the {\nr} region. The fitted nonlinear {\rts} agree very well with the experimental data.
As the linear formula is used to fit the {\rts} for the heavy mesons, $c_1$ which should be small becomes very large. It suggests that the linear formula is not a good match for the {\nr} energy region.

For the heavy mesons especially for the bottomonia, $\xi1$ [Eq. (\ref{delquant})] calculated by the nonlinear formula and $\xi2$ by the linear formula are consistent with each other and obey the constraints in (\ref{quantdelt}), see Tables \ref{tab:botenergy}, \ref{tab:radc} and \ref{tab:orbc}.
As shown in Table \ref{tab:botenergy}, $\zeta\ll1$ and $\xi\ll2$ are in agreement. It is expected that all bottomonia including the observed states and the states observed in the future will be in the {\nr} region. The predicted masses of the highly excited states of the bottomonia are also listed in Table \ref{tab:botenergy}.

\subsection{{\rts} for the heavy-light mesons}\label{subsec:rthlm}
For both the nonlinear fits [Eqs. (\ref{nlrts}) and (\ref{reghlf})] and the linear fits [Eq. (\ref{reglin})], all states on the {\rts} are used. Only the {\rts} with three or more points are presented.

\begin{figure*}[!phtb]
\centering
\subfigure[]{\label{fig:subfigure:bs}\includegraphics[scale=0.6]{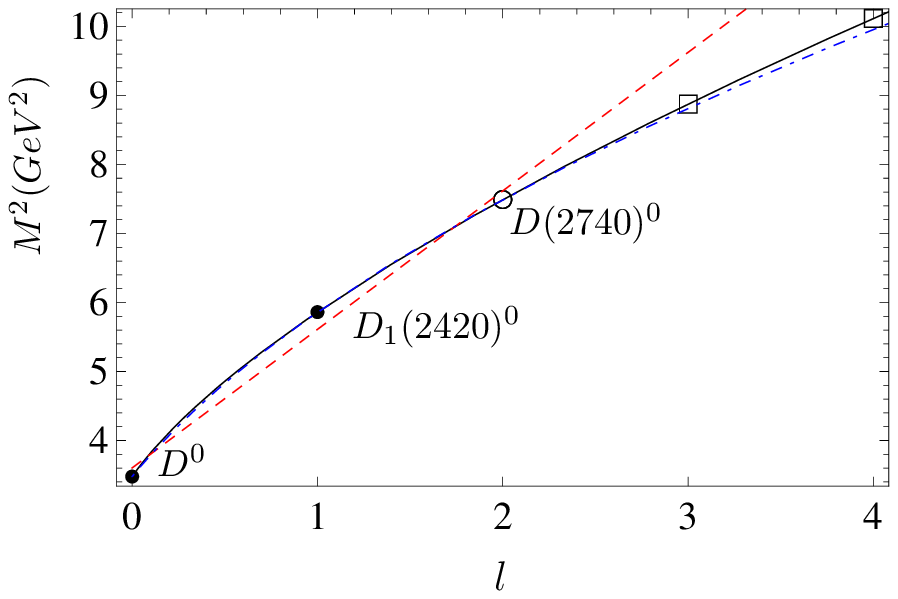}}
\subfigure[]{\label{fig:subfigure:bp}\includegraphics[scale=0.6]{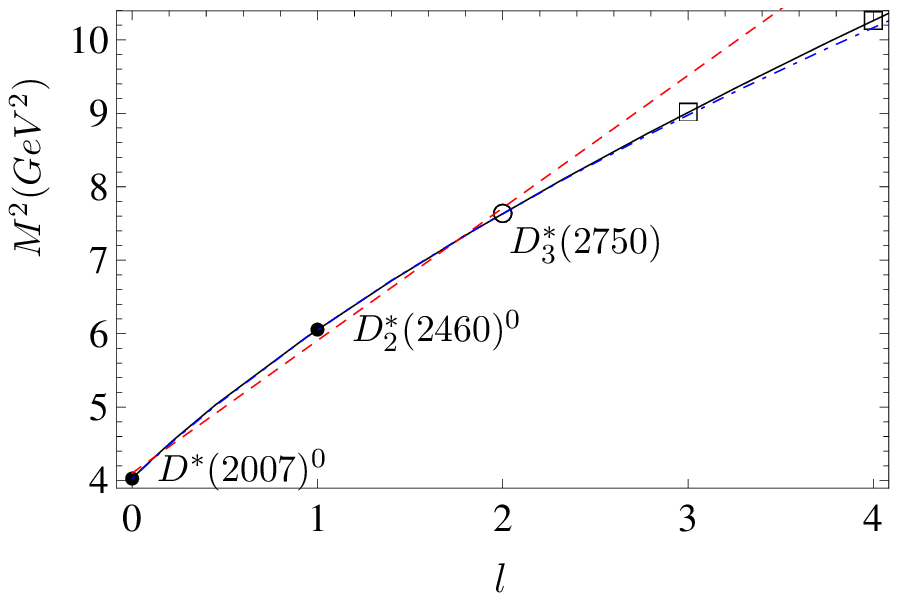}}
\subfigure[]{\label{fig:subfigure:bp}\includegraphics[scale=0.6]{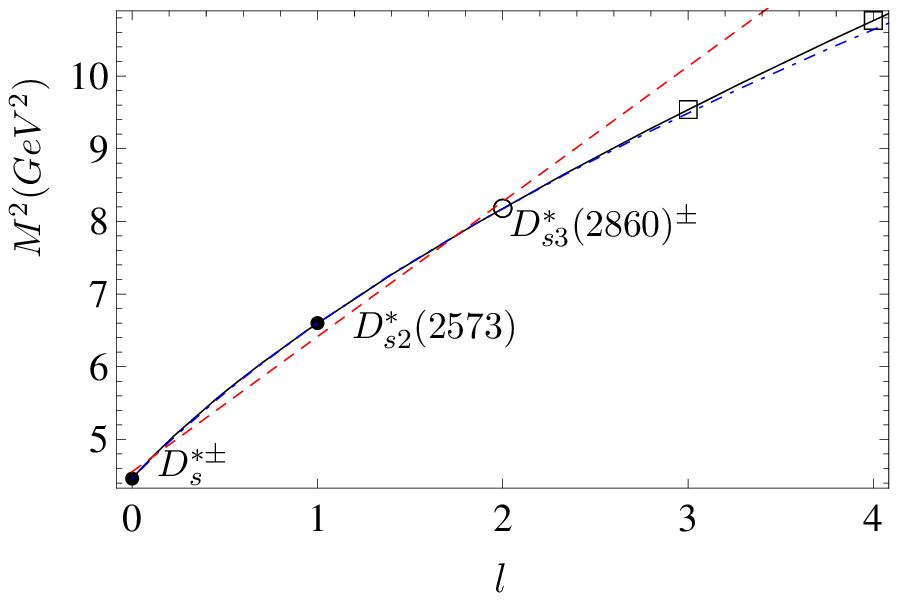}}
\caption{The {\rts} for the heavy-light mesons. The linear formula (\ref{reglin}) (the red dashed line), and the nonlinear formulas in Eq. (\ref{nlrts}) (the black line) and in Eq. (\ref{reghlf}) (the blue dot-dashed line) are employed. $M^2=4.55\sqrt{l+0.48}+0.32$ for the $D^0$ {\rt}, $M^2=5.10\sqrt{l+1.12}-1.37$ for the $D^{\ast}$ {\rt} and $M^2=4.74\sqrt{l+0.78}+0.27$ for the $D_s^{\ast}$ {\rt}. Other fitted {\rts} are listed in Table \ref{tab:orbc}.}\label{fig:rthl}
\end{figure*}

The quantity $\zeta{\sim}10^{-1}$ for the heavy-light mesons, see Table \ref{tab:orbmesons} and \ref{subsec:energy}. The {\rts} for the heavy-light mesons are nonlinear.
As the nonlinear formula (\ref{reghlf}), which is derived in case of the ideal heavy-light systems, is employed to fit the {\rts} for the heavy-light mesons, the fitted parameter $c_1$ is small or negative, see Fig. \ref{fig:rthl}. It disagrees with the constraints in Eq. (\ref{quantdelt}). As the nonlinear formula (\ref{nlrts}) is applied, $\xi1$ is greater than one which does not obey the constraints in (\ref{quantdelt}). $\xi2$ calculated by the fitted linear formula in (\ref{reglin}) neither obeys the constraints.
Moreover, $\xi1$ calculated by the nonlinear form and $\xi2$ by the linear form contradict, see Tables \ref{tab:radc} and \ref{tab:orbc}. All these clues indicate that the heavy-light mesons are not the ideal heavy-light systems, not in the {\nr} region nor in the {\ur} region. However, they can be regarded brute-forcely as the ideal heavy-light systems or as being in the intermediate energy region,

\subsection{{\rts} for the light mesons}\label{subsec:rtlm}
For the nonlinear fit, all states on the {\rts} are used. For the linear fit, the last four states on a {\rt} are used if the data points on the {\rt} are equal to or greater than four.

\begin{figure*}[!phtb]
\centering
\subfigure[]{\label{fig:piorb}\includegraphics[scale=0.6]{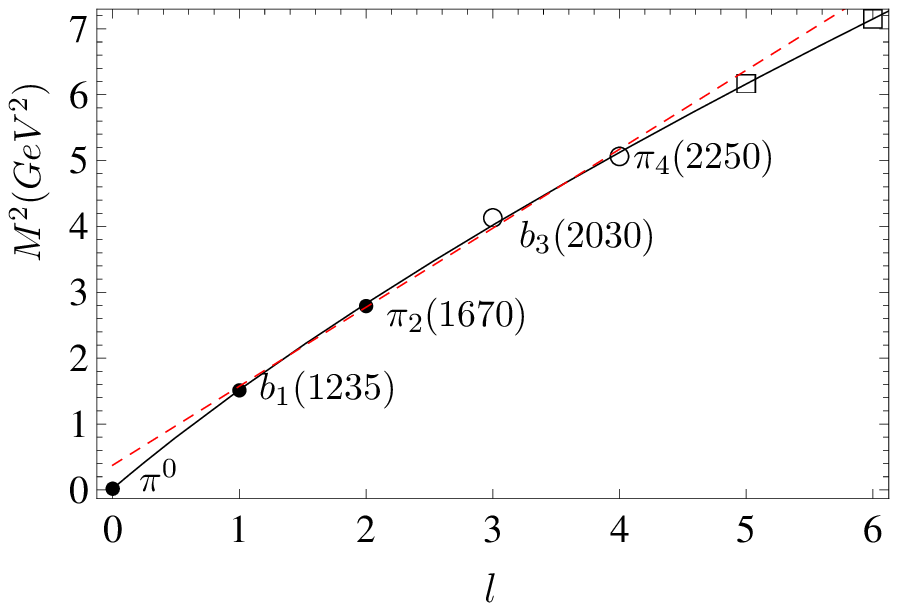}}
\subfigure[]{\includegraphics[scale=0.6]{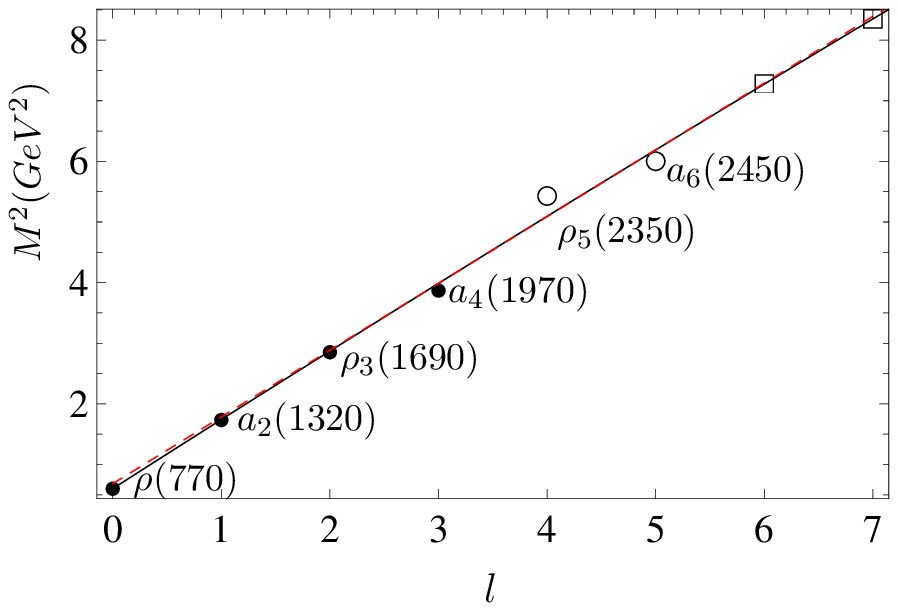}}
\subfigure[]{\includegraphics[scale=0.6]{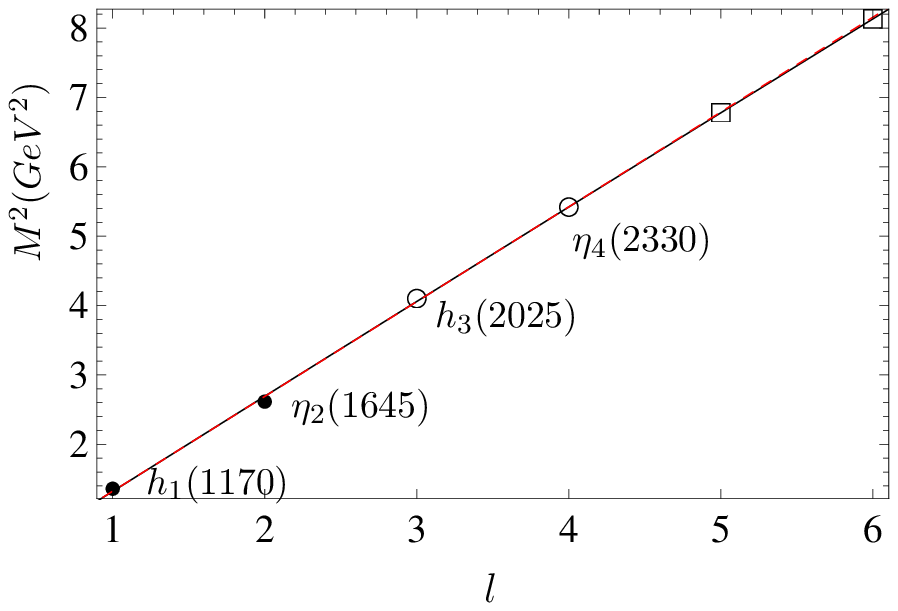}}
\subfigure[]{\includegraphics[scale=0.6]{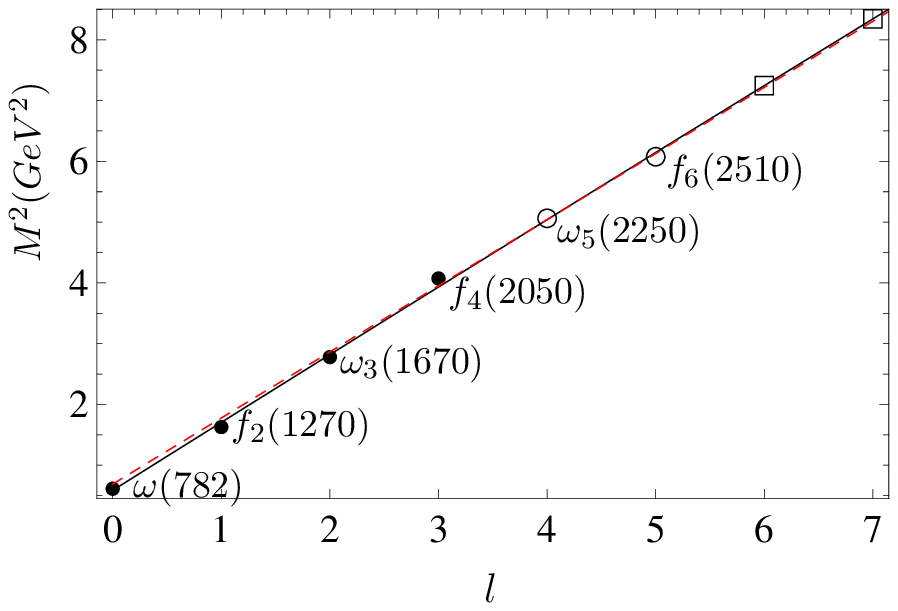}}
\subfigure[]{\includegraphics[scale=0.6]{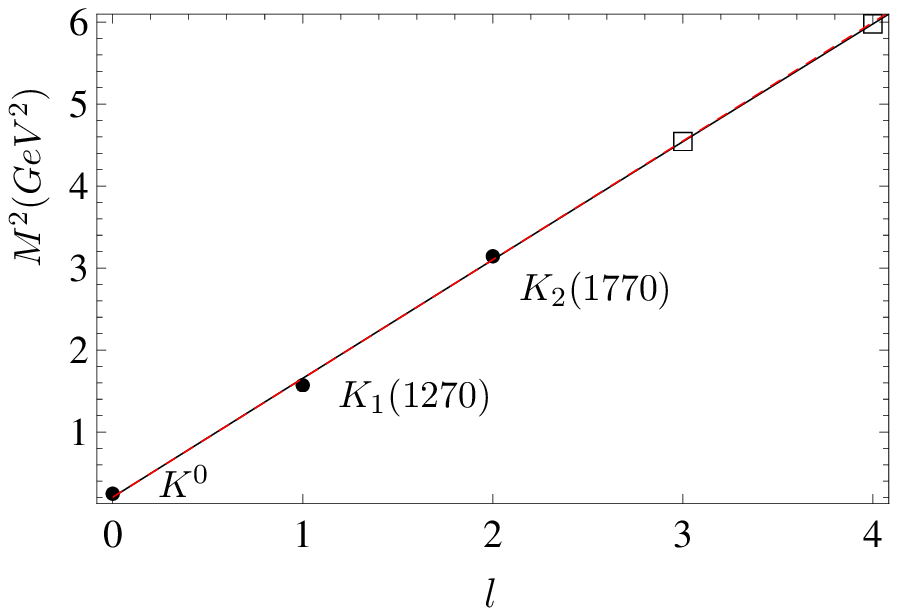}}
\subfigure[]{\includegraphics[scale=0.6]{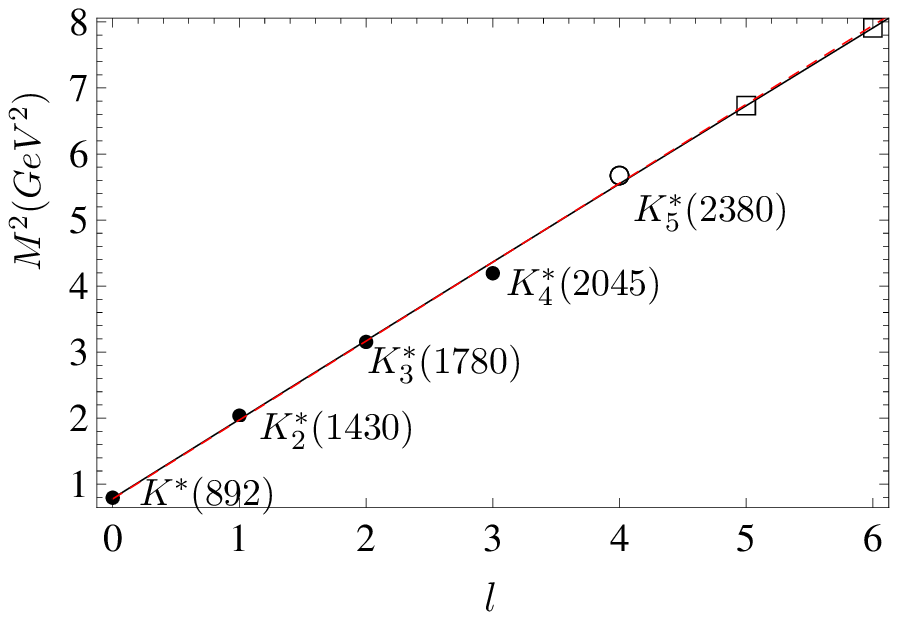}}
\subfigure[]{\includegraphics[scale=0.6]{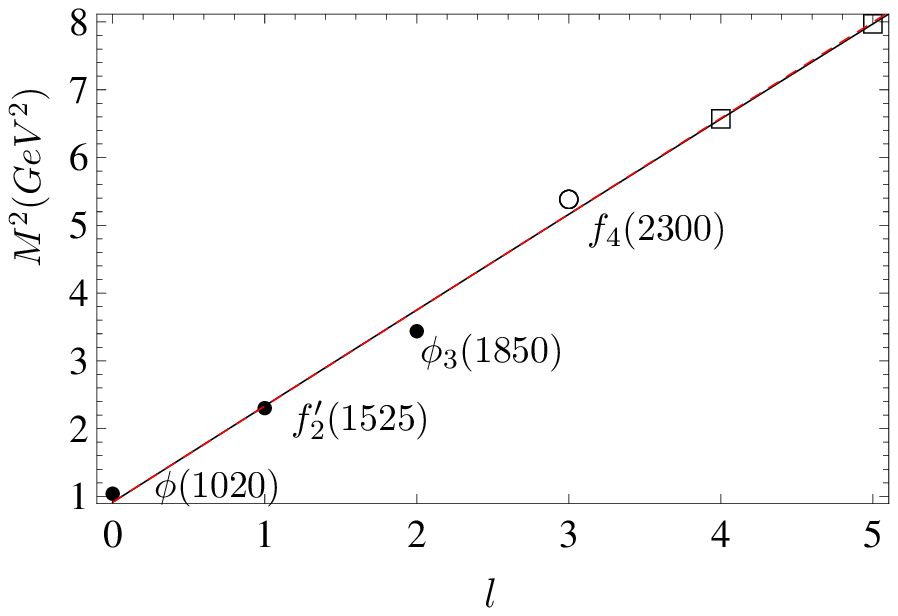}}
\subfigure[]{\label{fig:pir}\includegraphics[scale=0.6]{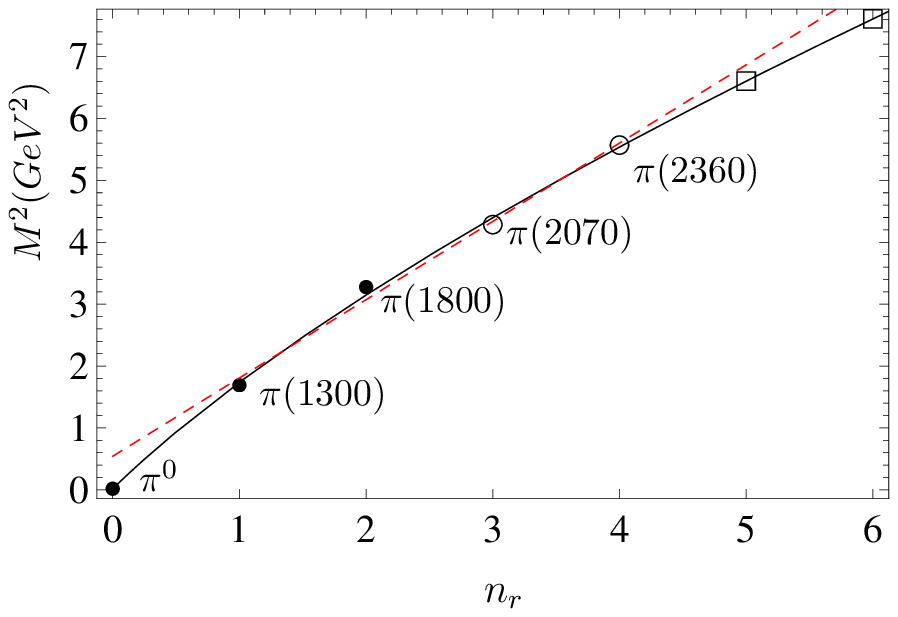}}
\subfigure[]{\includegraphics[scale=0.6]{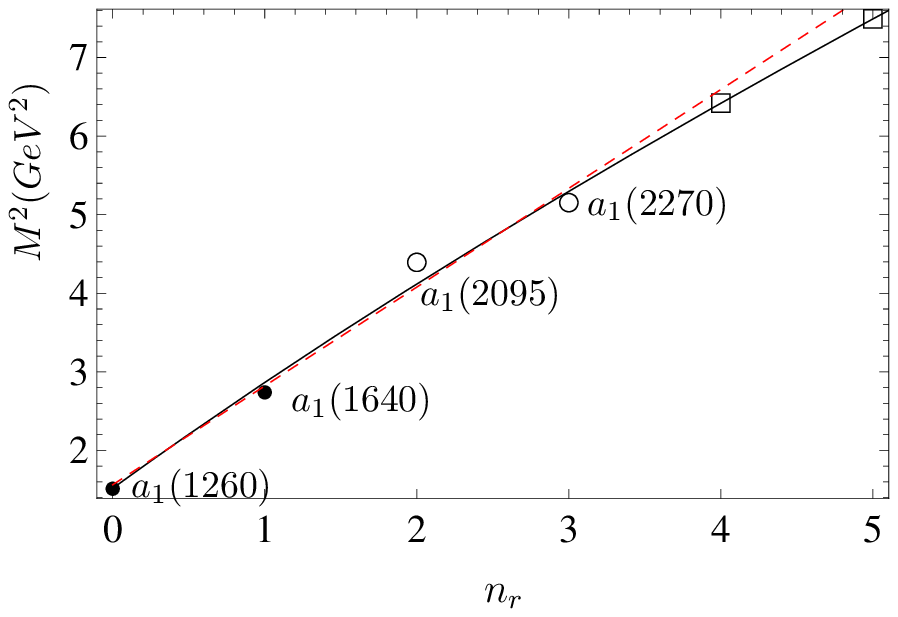}}
\subfigure[]{\includegraphics[scale=0.6]{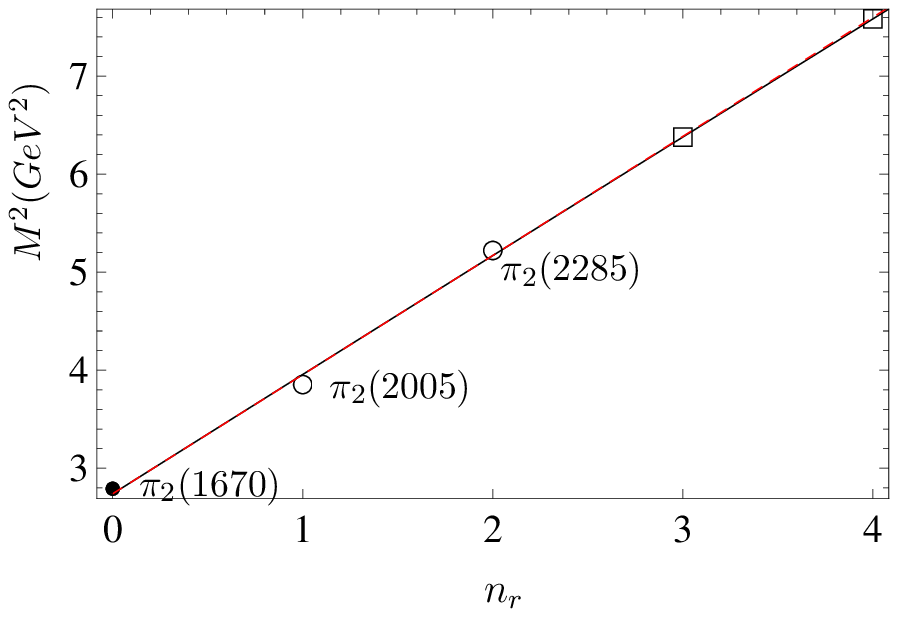}}
\subfigure[]{\includegraphics[scale=0.6]{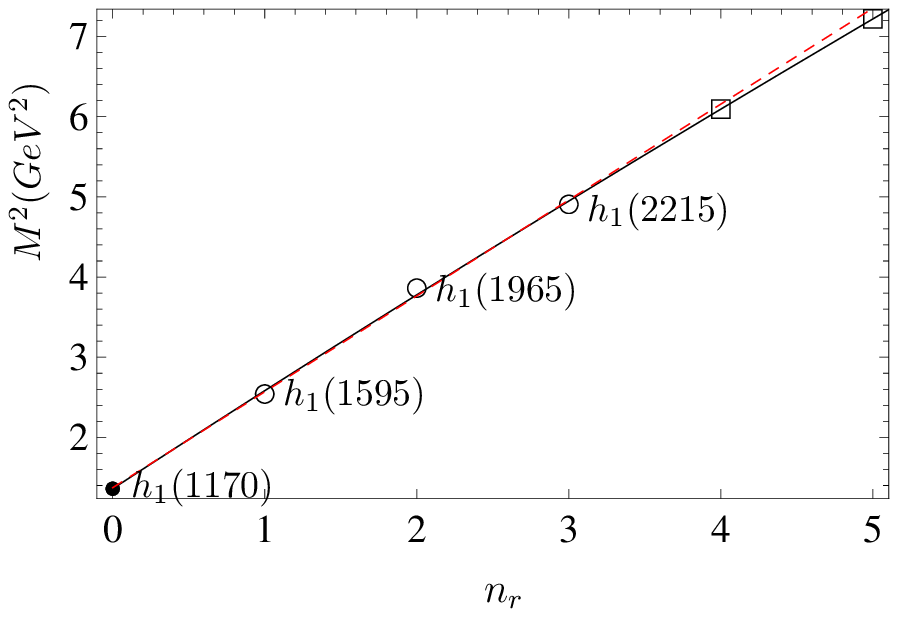}}
\subfigure[]{\includegraphics[scale=0.6]{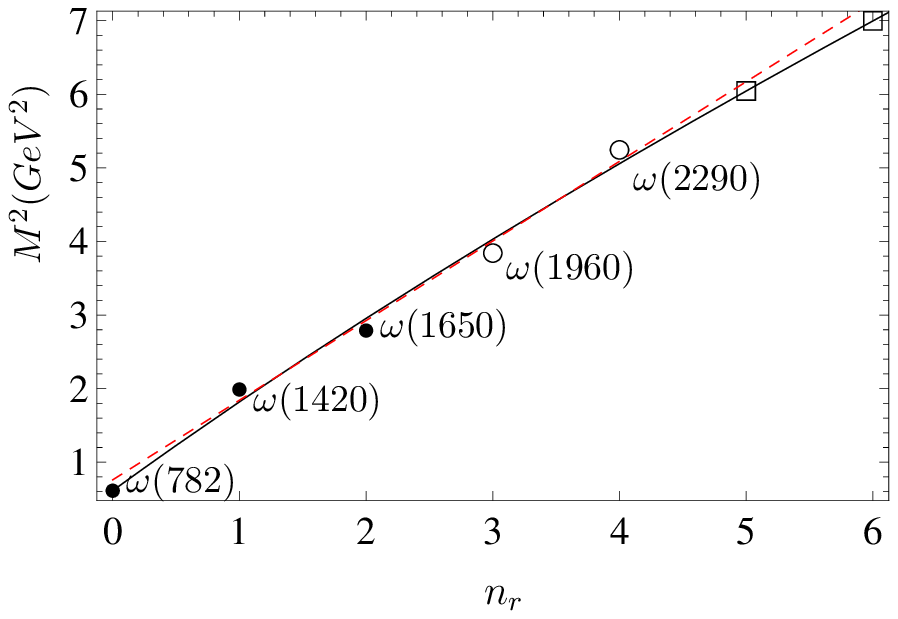}}
\subfigure[]{\includegraphics[scale=0.6]{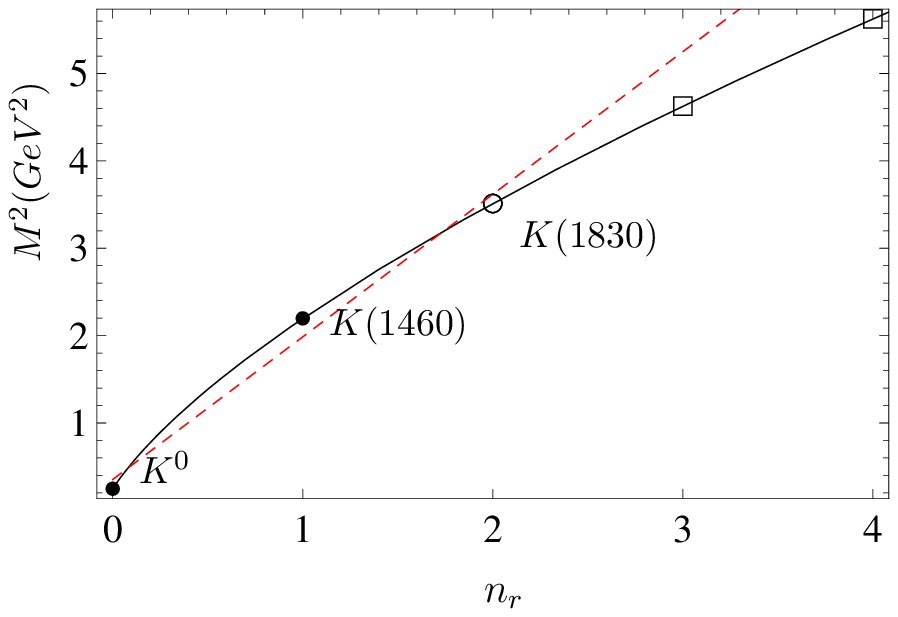}}
\subfigure[]{\includegraphics[scale=0.6]{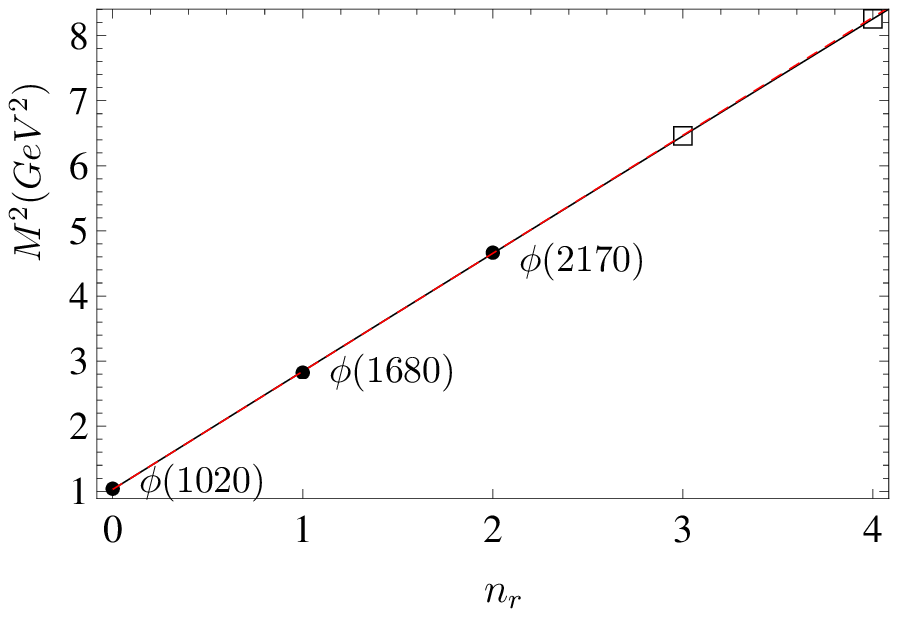}}
\caption{Same as Fig. \ref{fig:rthv} except for the light mesons.}\label{fig:rtlight}
\end{figure*}

As shown in Tables \ref{tab:radc} and \ref{tab:orbc}, $\xi$ calculated by the linear formula is much greater than $0.5$ while $\xi$ calculated by the nonlinear formula is negative. These results are consistent and show that the light mesons are in the relativistic region. ($c'$ and $\xi2$ for the linear $\eta/h$ {\rt} are negative, which can be adjusted if the first data point is included.)

The light mesons are the relativistic systems, see Tables \ref{tab:orbmesons} and \ref{tab:radmesons}. The {\rts} for the light mesons can be well described by the linear formula, see Fig. \ref{fig:rtlight}.
As applying the nonlinear formula appropriate for the {\nr} region to fit the {\rts} for the light mesons, $c_1$, which should be positive according to Eq. (\ref{quantdelt}), becomes negative, see Tables \ref{tab:radc} and \ref{tab:orbc}.
$c_0$ becomes very large and $c_0=100$ in this work. $\beta_x$ which should be small becomes large, see Tables \ref{tab:radc}, \ref{tab:orbc}, Eqs. (\ref{nrvalue}) and (\ref{regc1}). These imply that the nonlinear formula (\ref{nlrts}) does not match the {\ur} energy region. However, as shown in Fig. \ref{fig:rtlight}, the nonlinear formula (\ref{nlrts}) can give good extrapolated predictions because (\ref{nlrts}) is appropriate mathematically, see \ref{subsec:xidisc}.

According to Tables \ref{tab:orbmesons}, \ref{tab:radmesons}, \ref{tab:radc} and \ref{tab:orbc}, $\pi$ mesons are in the relativistic region. However, the {\rts} for $\pi$ are nonlinear, see Figs. \ref{fig:piorb} and \ref{fig:pir}. The cause of the nonlinearity remains unclear. It maybe is a coincidence that the {\rts} for $\pi$ can be well described by the nonlinear formula (\ref{nlrts}).

\subsection{Discussions on $\xi$}

Both $\zeta$ [Eq. (\ref{fzeta})] and $\xi$ [Eq. (\ref{delquant})] can measure the relation between the interaction energy and the masses of constituents.
$\zeta$ can give a clear classification of the energy regions while there exist ambiguity sometimes for $\xi$, see \ref{subsec:xidisc} and \ref{subsec:energy}.
Calculating the quantity $\zeta$ needs the masses of the constituents which vary with models while calculating $\xi$ does not need the masses.

For an unknown {\rt}, it is better to employ the linear formula and the nonlinear formula to calculate the quantity $\xi$ to check the energy region. As $\xi\ll2$, the mesons will be {\nr} and the nonlinear formula (\ref{nlrts}) is appropriate. As $\xi\gg0.5$ or $\xi<0$, the mesons will be relativistic and the linear formula is good. If $\xi$ calculated by two different formulas are in contradiction, the mesons will be regarded rudely as being in the intermediate region,  see \ref{subsec:rthlm}, Tables \ref{tab:radc} and \ref{tab:orbc}.

\section{Dependence of $\beta$ on mass and the string tension $\sigma$}\label{sec:beta}

The slope $\beta$ in the {\rts} is an important quantity \cite{Basdevant:1984rk}.
In the {\ur} limit, masses of the constituents are supposed to approach zero. It agrees with the well-known knowledge that the {\rts} for the light mesons is approximately linear and the slope depends only on the string tension $\sigma$, see Eqs. (\ref{reglin}) and (\ref{urvalue}). In the {\nr} limit, the {\rts} is significantly nonlinear and $\beta$ depends not only on the string tension but also on the masses of the constituents, see Eqs. (\ref{nlrts}) and (\ref{nrvalue}).
In the intermediate region, it is expected that $\beta$ depends likewise both on the masses of constituents and on the string tension, but the dependence on the masses are expected to weaken as the masses decrease.

Use a linear formula
\bea\label{mleq}
M^2=\beta' x+c'_1\; (x=n_r,\,l)
\eea
to fit a nonlinear {\rt} which can be described by
\bea\label{mnleq}
M^2=\beta_1 (x+c_0)^{\nu}+c_1\; \text{or}\; M^2=\beta_{2a}x+\beta_{2b}x^{\nu}+c_2,
\eea
where $\beta'$ and $\beta_{2a}$ are independent of the masses of the constituents while $\beta_1$ and $\beta_{2b}$ depend on the masses. By differentiating Eqs. (\ref{mleq}) and (\ref{mnleq}), we have
\bea\label{mbeta}
\beta'=\nu{\beta_1}(x_0+c_0)^{\nu-1}\; \text{or}\; \beta'=\beta_{2a}+\nu\beta_{2b}x^{\nu-1},
\eea
where $x_0$ is a point on the {\rt}.

In case of the linear fit, Eq. (\ref{mleq}) is employed. The fitted slopes
are almost a constant for the light mesons. They increase for the heavy-light mesons and become very large for the heavy mesons.
It is in agreement with Refs. \cite{Nielsen:2018ytt,Ebert:2009ub,Ebert:2009ua,Ebert:2011jc,Sonnenschein:2014jwa,
Abreu:2020ttf,Abreu:2020wio,Kher:2017mky}.
The fitted slopes of the radial trajectories $\beta'_{n_r}$ are about 1.2 for the light unflavored mesons, become 1.63 for the $K$ trajectory, 3.26 for the $\psi$ trajectory and 4.79 for the $\Upsilon$ trajectory. $\beta'_l$ are about 1.1 for the light unflavored mesons, $1.45$ for the $K$ trajectory, $1.80$ for the $D^\ast$ trajectory, 3.06 for the $\psi$ trajectory and 8.76 for the $\Upsilon$ trajectory, see Tables \ref{tab:radc} and \ref{tab:orbc}.
Eq. (\ref{mbeta}) explains why the $\beta'$ of (\ref{mleq}) increases with the masses of the constituents. For example, for the $\psi/\chi_c$ trajectory $\beta_l=3.53$, $c_0=0.08$ and $x_0=0.375$ give $\beta'_l=3.06$ according to Eq. (\ref{mbeta}), which is in agreement with the fitted value.

In case of the nonlinear fit, the first formula in (\ref{mnleq}) with $\nu=2/3$ is applied, which is appropriate for the heavy mesons, see Fig. \ref{fig:rthv}. The fitted slopes $\beta_l$ and $\beta_{n_r}$ are in accordance with the theoretical predictions, see Ref. \cite{Chen:2018hnx}. As discussed in the subsection \ref{subsec:xidisc}, the first formula in (\ref{mnleq}) can also be applied to fit the {\rts} for the heavy-light mesons and for the light mesons, see Figs. \ref{fig:rthl} and \ref{fig:rtlight}.
According to Eq. (\ref{nlapp}), there is a relation between the slopes of a nonlinear fit and the slopes of a linear fit. For example, for the $\rho/a$ trajectory $\beta'_{l}=1.15$ calculated by using Eq. (\ref{nlapp}) is in agreement with the fitted value $\beta'_l=1.10$.

\section{Conclusions}\label{sec:con}
In this work, we investigate the structure of the meson Regge trajectories based on the {\qsse}. The form of the meson {\rts} is complicated and depends on the energy region. In the {\nr} limit, the approximated form of the {\rts} is $M^2=\beta_x(x+c_0)^{2/3}+c_1$ [Eq. (\ref{nlrts})]. In the {\ur} limit, it is well known that the {\rts} can be well described by the linear formula $M^2=\beta_{x}x+c_1$ [Eq. (\ref{reglin})]. In the intermediate energy region, the simple form of the {\rts} remains unclear and is expected to be nonlinear.
We show that the Regge trajectories obtained from different approaches are consistent with each other in the {\nr} limit and in the {\ur} limit.

By employing the nonlinear formulas and the linear formula, the Regge trajectories for different mesons are given. As a {\rt} formula is unsuitable for the energy region, the fitted parameters neither have explicit physical meanings nor obey the constraints [Eq. (\ref{quantdelt})], however, the fitted {\rt} can give the satisfactory extrapolated predictions if the employed formula is appropriate mathematically. Using a linear formula to fit the {\rts} for the heavy mesons, $\beta_x$ and $c_1$ become very large. Conversely, using a nonlinear formula to fit the light mesons, $c_1$ becomes negative.

Moreover, the slopes of the fitted linear formula will increase for the heavy-light mesons and become very large for the heavy mesons.
We show that the masses of the constituents will come into the slope and explain why the slopes of the fitted linear {\rts} vary with the masses of the constituents as the linear formula is used to fit the {\rts} for the heavy-light mesons and for the heavy mesons, see Eq. (\ref{mbeta}).

\flushleft
{\bf Acknowledgements}
We are very grateful to the anonymous referees for the valuable comments and suggestions.
This work is supported by the Natural Science Foundation of Shanxi Province of China under Grant no. 201901D111289.

\flushleft
{\bf Data Avilability Statement} This manuscript has no associated data or the data will not be deposited. [Authors' comment: All data are included in the paper.]

\begin{table*}[!phtb]
\centering
\caption{The radial {\rts} fitted by the nonlinear formula $M^2=\beta_{n_r}(n_r+c_0)^{2/3}+c_1$ (Fit1) and by the linear formula $M^2=\beta^{\prime}_{n_r} n_r+c'$, $c'=\beta^{\prime}_{l}l + c'_1$ (Fit2). $\beta^{\prime}_{l}{\approx}1.1$ for $a_1$, $\pi_2$ and $h_1$. $\beta^{\prime}_{l}\approx5.1$ for $\chi_{b1}$ and $\chi_{b2}$ by fitting the data in Ref. \cite{Ebert:2011jc}. For the nonlinear fit, all points are used. For the linear fit, only the last four points are used if the points on the {\rt} are more than four. For the nonlinear fit, the effect of $l$ is absorbed into $c_0$, see Eq. (\ref{nrconst}). For the linear fit, the terms containing $n_r$ and $l$ should be written explicitly, see Eq. (\ref{urconst}). $\dag$ denotes that $\xi1$ reads $0.48$ by using the global fit \cite{Chen:2018hnx}. }\label{tab:radc}
\begin{tabular*}{0.9\textwidth}{@{\extracolsep{\fill}}ccccc@{}}
\hline\hline
 Traj.         & Fit1                               & Fit2                  & $\xi1$   & $\xi2$   \\
               & $(\beta_{n_r},c_0,c_1)$                 &  $(\beta'_{n_r},c')$       & $n_r=1$   &  $n_r=1$\\
\hline
  $\pi$         & $(2.78,\;0.80,\;-2.38)$      &$ (1.26,\;0.54)$    & $-$1.73   & 2.33    \\
  $a_1$         & $(3.23,\;3.59,\;-6.05)$      & $(1.26,\;1.56)$    & $-$1.47   & 5.13       \\
  $\pi_{2}$     & $(8.49,\;100,\;-180.)$      & $(1.22,\;2.74)$     & $-$1.02   & 6.33  \\
  $h_{1}$       & $(4.42,\;13.5,\;-23.7)$      & $(1.20,\;1.37)$     & $-$1.11  & 8.52 \\
  $\omega$      & $(3.06,\;4.33,\;-7.52)$      & $(1.08,\;0.76)$     & $-$1.24   &1.42 \\
  $K$           & $(2.29,\;0.10,\;-0.25)$      & $(1.63,\;0.35)$     &$-$9.76     &4.66 \\
  $\phi$        & $(12.7,\;100,\;-272.)$      & $(1.81,\;1.03)$     & $-$1.01    & 1.76\\
$D$\cite{Chen:2018nnr} &$( 4.05,\; 0.25,\; 1.85)$ & $( 2.68,\; 3.62)$  &\hspace{1.1mm} 2.54    & 0.74 \\
  $\psi$        & $(6.19,\;0.77,\;4.39)$      & $(3.26,\;9.87)$     & \hspace{3.8mm}$2.06^{\dag}$    &0.33 \\
  $\Upsilon$    & $(11.0,\;0.0,\;89.5)$       & $(4.79,\;97.9)$     & \hspace{1.1mm}  0.12    &0.05 \\
  $\chi_{b1}$   & $(9.67,\;0.27,\;93.8)$      & $(6.33,\;98.2)$     &\hspace{1.1mm}  0.12   & 0.12\\
  $\chi_{b2}$   & $(9.59,\;0.28,\;94.1)$      & $(6.25,\;98.6)$     &\hspace{1.1mm}  0.12    & 0.12\\
\hline\hline
\end{tabular*}
\end{table*}

\begin{table*}[!phtb]
\centering
\caption{The orbital {\rts} fitted by the nonlinear formula $M^2=\beta_l(l+c_0)^{2/3}+c_1$ (Fit1) and by the linear formula $M^2=\beta^{\prime}_{l} l+c'$ (Fit2), where $c'=\beta^{\prime}_{n_r}n_r + c'_1$, $n_r=0$. For the nonlinear fit, all points are used. For the linear fit, only the last four points are used if the points on the {\rt} are more than four. For the nonlinear fit, the effect of $n_r$ is absorbed into $c_0$, see Eq. (\ref{nrconst}). For the linear fit, the terms containing $n_r$ and $l$ should be written explicitly, see Eq. (\ref{urconst}).}\label{tab:orbc}
\begin{tabular*}{\textwidth}{@{\extracolsep{\fill}}cccrr@{}}
\hline\hline
 Traj.         & Fit1                               & Fit2                  & $\xi1$   & $\xi2$   \\
               & $(\beta_{l},c_0,c_1)$                 &  $(\beta'_{l},c')$       & $l=1$   &  $l=1$\\
\hline
  $\pi/b$         & $(2.83,\; 1.52,\; -3.73)$        &$ (1.20,\; 0.38)$   & $-$1.41     &  3.16   \\
  $\rho/a$         & $(5.36,\; 30.3,\; -51.5)$       &$ ( 1.10,\; 0.68)$   & $-$1.03     & 1.62     \\
  $\eta/h$         & $( 9.59,\; 100,\; -207)$       &$ ( 1.37,\; -0.04)$   & $-$1.00     & $-$34.3     \\
  $\omega/f$         & $( 7.81,\; 100,\; -168)$      &$ ( 1.09,\; 0.69)$   & $-$1.01     & 1.58    \\
  $K$             & $( 10.1,\; 100,\; -218)$         &$ ( 1.45,\; 0.21)$   & $-$1.00     & 6.90     \\
  $K^{\ast}$         & $( 8.35,\; 100,\; -179)$      &$ ( 1.20,\; 0.78)$   & $-$1.01     & 1.54     \\
  $\phi/f'$         & $( 9.91,\; 100,\; -213)$       &$ ( 1.42,\; 0.92)$   & $-$1.01     & 1.54     \\
  $D$              & $( 2.85,\; 0.12,\; 2.78)$       &$ ( 2.01,\; 3.60)$   & 1.11     &   0.56   \\
  $D^{\ast}$         & $( 2.97,\; 0.49,\; 2.18)$      &$ ( 1.80,\; 4.10)$   & 1.78     &  0.44    \\
   $D^{\ast}_s$      & $( 2.86,\; 0.29,\; 3.21)$      &$( 1.86,\; 4.55)$   &  1.06    &   0.41   \\
   $\psi/\chi_c$\cite{Chen:2018hnx}   & $(3.53,\; 0.08,\; 8.92)$ &$(3.06,\; 9.59)$ & 0.42 & 0.32 \\
   $\Upsilon/\chi_b$\cite{Chen:2018hnx} &$( 9.27,\; 0.02,\; 88.9)$ &$( 8.76,\; 88.5)$ &0.11  &0.10  \\
\hline\hline
\end{tabular*}
\end{table*}


\end{document}